\documentclass[12pt]{article}
\usepackage{fullpage,setspace}
\usepackage{amssymb, amsthm, amsmath}
\usepackage{graphicx}
\usepackage[authoryear]{natbib}
\usepackage{bm}
\usepackage{lineno}
\usepackage{times}
\usepackage{xcolor}
\usepackage{algorithm}
\usepackage{algpseudocode}
\usepackage{pdflscape}
\usepackage{multirow}
\usepackage{booktabs}
\usepackage{caption}
\usepackage{subcaption}

\newcommand{\bb}{ \mbox{\bf b}}

\newcommand{\bx}{ \mbox{\bf x}}
\newcommand{\bX}{ \mbox{\bf X}}

\newcommand{\bZ}{ \mbox{\bf Z}}

\newcommand{\bY}{ \mbox{\bf Y}}

\newcommand{\bs}{ \mbox{\bf s}}

\newcommand{\bU}{ \mbox{\bf U}}

\newcommand{\bW}{ \mbox{\bf W}}

\newcommand{\iid}{\stackrel{iid}{\sim}}

\newcommand{\beq}{ \begin{equation}}
\newcommand{\eeq}{ \end{equation}}
\newcommand{\beqn}{ \begin{eqnarray}}
\newcommand{\eeqn}{ \end{eqnarray}}

\usepackage{soul}

\usepackage{caption}
\begin{document}  %\linenumbers

\begin{center}
{\Large Density correction for multivariate spatial fields of global climate model output using deep learning}\\\vspace{6pt}
{\large Reetam Majumder\footnote[1]{Department of Mathematical Sciences, University of Arkansas}, Shiqi Fang\footnote[2]{Department of Civil, Construction, and Environmental Engineering, North Carolina State University}, A. Sankarasubramanian\textsuperscript{2},\\ Emily C. Hector\footnote[3]{Department of Statistics, North Carolina State University}, Brian J. Reich\textsuperscript{3}}\\
%\today
\end{center}

\begin{abstract}\begin{singlespace}\noindent
Global Climate Models (GCMs) are numerical models that simulate complex physical processes within the Earth's climate system and are essential for understanding and predicting climate change. However, GCMs suffer from systemic biases due to simplifications made to the underlying physical processes. GCM output therefore needs to be bias corrected before it can be used for future climate projections. Most common bias correction methods, however, cannot preserve spatial, temporal, or inter-variable dependencies. We propose a new semi-parametric estimation of conditional densities (SPECD) approach for density correction of the joint distribution of daily precipitation and maximum temperature data obtained from gridded GCM spatial fields. The Vecchia approximation is employed to preserve dependencies in the observed field during the density correction process, which is carried out using semi-parametric quantile regression. 
The ability to calibrate joint distributions of GCM projections has potential advantages not only in estimating extremes, but also in better estimating compound hazards, like heat waves and drought, under potential climate change. Illustration on historical data from 1951--2014 over two $5\times 5$ spatial grids in the US indicate that SPECD can preserve key marginal and joint distribution properties of precipitation and maximum temperature, and predictions obtained using SPECD are better calibrated compared to predictions using asynchronous quantile mapping and canonical correlation analysis, two commonly used bias correction approaches.
\vspace{12pt} \end{singlespace}\end{abstract}

%\newpage

\section{Introduction}\label{s:intro}
Global Climate Models (GCMs) are %physical representations of complex interactions between ocean, atmosphere and land surface. They are 
numerical models to explain complex interactions between the ocean, atmosphere, and land surface. 
GCMs are used to study long-term climate trends and to provide projections based on possible future scenarios associated with, e.g., greenhouse gas and aerosol emissions, land use, and population growth \citep{Gettelman2016-jk}. The latest generation of GCMs are collectively known as the Climate Model Intercomparison Project version 6 \citep[CMIP6;][]{Eyring2016}. It contains the outputs, primarily ocean, atmospheric, and land surface conditions, as ensembles at a coarser spatial scale for further analyzing impacts on various sectors. For instance, GCM forcings (typically precipitation and temperature) are used to assess for both surface water and groundwater availability over planning periods \citep{singh2015,seo2016}. To develop these assessments of future water availability, GCM projections, particularly for precipitation and temperature, are forced through hydrologic models, which are subsequently used as an input into a reservoir model \citep{singh2015,seo2016}. Projected inflow into reservoirs are typically used for allocating water for multipurpose uses that includes municipal, irrigation, hydropower and flood control. GCM projections for other variables, particularly temperature and humidity, are also critical for quantifying energy demand \citep{ESHRAGHI2021121273}. 
%However, spatial mismatch between the forcings of GCMs and hydrological models and other sectoral models (e.g., energy demand prediction), GCM projections typically require bias corrections and downscaling for application into climate impact studies.  

Recent computational advances and parameterization of physical processes (e.g., cloud physics) have improved the resolution and accuracy of GCM projections under potential emission scenarios \citep{RamirezVillegas2013,strandberg2021}. 
Despite these advances, GCM projections have some key drawbacks which need to be addressed before they can be used in climate change impact studies. First, they have systematic biases due to model errors that arise from simplifications made to the underlying physical processes \citep{stevens2013,Du2022}. 
These biases lead to discrepancies between observed and model data resulting in an inability to estimate the observed statistics of climate variables \citep{Bhowmick2019}. 
%and render GCM output unusable at local \citep{TEUTSCHBEIN201212} and regional \citep{STRAFFELINI2023103647} scales. 
Such discrepancies between modeled and observed data also result in error propagation in sectoral analysis (e.g., water management) \citep[][]{singh2015,seo2016}. In addition, hydrologic projections developed using GCM projections may exhibit uncertainty due to these imprecise forcings \citep{bhowmick2017,SEO2019304} or from (long-term) scenarios  \citep{Lettenmaier1999,Hawkins2011}, adding challenges in sectoral application.
Second, the coarse resolution of GCMs often fails to fully resolve the effect of topography or sub-grid variability on hydroclimatic variables, especially in projecting precipitation. While some CMIP6 models have a 50 km spatial resolution, most are between 100--250 km. At these resolutions, topography, land use, and land cover are poorly represented; GCMs are therefore unable to accurately capture the land surface and atmospheric feedback that drive local hydroclimatic extremes \citep{Li-2011,OBrien2016}. These two problems are addressed by bias-correcting and downscaling the GCM output, respectively, carried out over the historical period where both observational and model data are available. The bias correction and downscaling models can then be used to obtain fine-scale, unbiased projections of future climate.

%The remainder of this manuscript focuses on the bias correction of daily precipitation (PRCP) and maximum temperature (TMAX) data based on CMIP6 GCMs. 
Traditional bias correction methods like asynchronous quantile mapping (QM) have been used to adjust the distribution of GCM data by mapping quantiles from the GCM to the corresponding quantiles in the observational record \citep{stoner2013,Thrasher2012,Maruan2013}. While effective for single-variable adjustments, QM often fails to account for inter-variable dependencies \citep{Bhowmick2019}. 
Multivariate bias correction methods aim to preserve the relationships between climatic variables \citep{bhowmick2017}. While multivariate QMs attempt to address this issue \citep{Pianni2012,Cannon2018}, QM is inappropriate for the bias correction of future projections as it distorts future climate change in a nonphysical way \citep[e.g.,][]{Maurer-2014}. Other commonly used methods such as bias corrected spatial downscaling \citep[BCSD,][]{wood2002}, bias corrected constructed analogues \citep[BCCA,][]{maurer2010}, and multivariate adapted constructed analogues \citep[MACA,][]{abatzoglou2012} also employ variants of QM as part of their framework. Multivariate methods like localized constructed analogs \citep[LOCA;][]{locav1,locav2} and asynchronous canonical correlation analysis \citep[CCA;][]{bhowmick2017} offer improvements by better capturing the temporal covariability in climate variables (e.g., cross-correlation), but do not explicitly consider spatial dependence during bias correction. Finally, recent deep learning-based methods, like DeepSD \citep{vandal2017}, leverage the power of deep learning algorithms to model complex, nonlinear relationships between climate variables and local physiographical information such as elevation, but have not been evaluated for the dependency between multiple climate variables, or the spatial dependence within each variable. 
Additionally, none of the aforementioned methods, with the exception of LOCA~v2 \citep{locav2}, are geared towards reproducing observed extreme quantiles of climate variables. %This limits their practical utility for studying climate extremes.
However, LOCA v2 also fails to preserve spatial dependency. We argue that bias correction should be viewed as a problem of \textit{density correction} - calibrating the spatiotemporal joint density of the historical GCM projections to the spatiotemporal joint density of the observed multivariate climate data. Subsequently, the `calibration' model will be run with future GCM projections to develop density-corrected (i.e., calibrated) GCM projections that remove the systematic biases in the GCM joint densities. Our effort here is to develop such a density correction approach by comparing the joint density fields of GCM output with its observed counterpart.

In this paper, we develop a new semi-parametric estimation of conditional densities (SPECD) approach for density correction of the joint distribution of daily precipitation (PRCP) and maximum temperature (TMAX) data. The joint distribution of the two variables over a spatial field is decomposed into a product of univariate conditional distributions using a Vecchia approximation \citep{vecchia1988estimation}, and each univariate density is estimated using semi-parametric quantile regression \citep[SPQR;][]{xu-reich-2021-biometrics}. SPQR uses neural networks to learn univariate conditional densities of random variables. The deep learning model underpinning SPQR ensures that nonlinear relationships are captured, and the Vecchia approximation ensures that spatial and cross correlations are preserved. We test our method with data over twenty five $100 \mbox{ km} \times 100 \mbox{ km}$ grid cells in the Southeast (SE) and Southwest (SW) US, depicted in Figure \ref{f:locations}. We use daily data from 1951--2000 to train our density correction model, and data from 2001--2014 to validate our results. Our results indicate that SPECD is able to preserve key marginal and joint distribution properties of TMAX and PRCP. Additionally, SPECD predictions are better calibrated over training and validation periods compared to QM and CCA for our spatial domain.

The proposed SPECD connects conditional density estimation \citep[CDE;][]{Hyndman1996,liracine2006} and bias correction in a unique manner, and in the process mitigates several drawbacks of commonly used bias correction methods. A key aspect of SPECD (described in further detail in Section \ref{s:mv}) is the estimation of the joint distribution of GCM and observational data within a single SPQR model, and thereafter using cumulative density function (CDF) and quantile function (QF) transforms to calibrate the model data. Similar CDF based approaches have been used for bias correcting future climate projections \citep[e.g.,][]{Michelangeli2009,Li2010}; however, these approaches estimate the CDF of the GCM and observed data independent of each other. 
SPECD also bears resemblance to normalizing flows \citep[NFs;][]{Rezende2015,Papamakarios2021,Kobyzev2021} and transport maps \citep[][]{Marzouk2016,Katzfuss02042024}. Both of these methods aim to transform a (usually complex) target distribution to a much simpler reference distribution (e.g., a standard Gaussian) through a series of invertible transformations; the transformations are referred to as flows and transport maps, respectively. Inverting the transformations allows for sampling from the target distribution as well as making predictions. Transport maps can be considered a special case of an NF \citep{Katzfuss02042024}. Similarly, SPECD parallels NFs - we can consider the GCM and observational data to be distributed according to the reference and target distributions respectively, with the flows being a sequence of CDFs and QFs. This also highlights the key difference between SPECD and NFs. While the reference distribution in NFs is assumed to be known, the joint distribution of the GCM data also needs to be learned alongside the joint distribution of the observational data. While this is a more complex task, SPECD's ability to estimate them simultaneously makes the calibration step of density correction quite straightforward.

The remainder of this paper is organized as follows. Section \ref{s:data-methods} introduces the data and methods for our study. Section \ref{s:sim} studies the performance of SPECD using a numerical study. Section \ref{s:app} contains a case study for density correction of climate model data at two different regions of the US. Section \ref{s:discussion} concludes.

\section{Data and Methods}\label{s:data-methods}
\subsection{Overview}
Section \ref{s:data} describes the different datasets used in our study, including resolution and pre-processing used to align the different datasets on a common spatial grid. Section \ref{s:mv} introduces the SPECD approach for density correction of TMAX (Kelvin) and PRCP (mm) at a single grid cell. Sections \ref{s:mvs} extends SPECD to the spatial setting. Section \ref{s:spqr} outlines SPQR, the conditional density estimator that is used for SPECD. Finally, Section \ref{s:gof} describes the metrics used to assess the quality of the calibrated data. 

\subsection{Temperature and precipitation datasets}\label{s:data}

\begin{figure}
    \centering
    \includegraphics[width=0.3\textwidth]{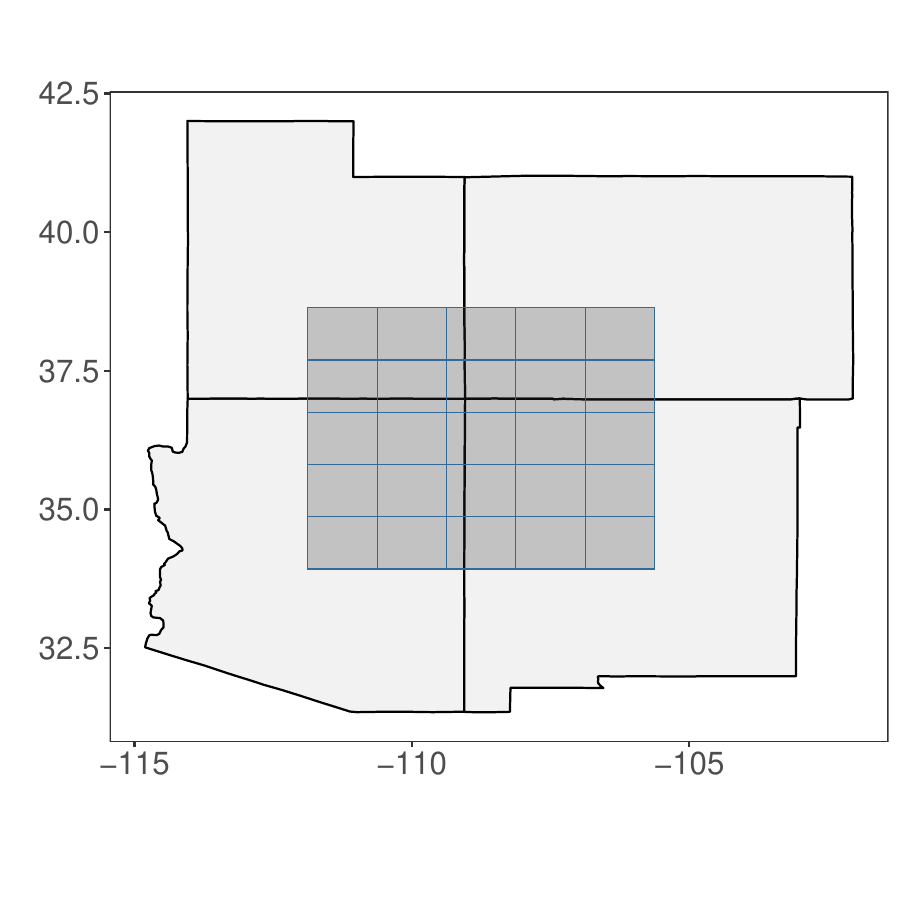}
    \includegraphics[width=0.32\textwidth]{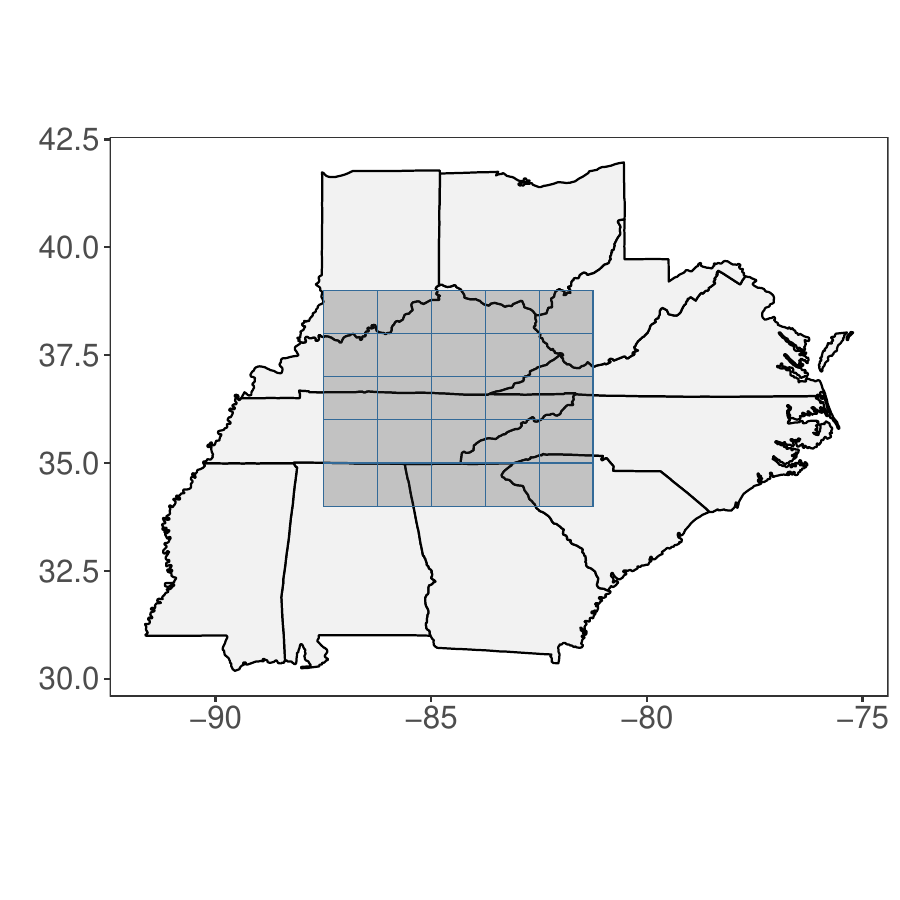} 
       \caption{Locations in the Southwest (left) and Southeast (right) regions of the US, which form the basis of our bias-correction case study.}
    \label{f:locations}
\end{figure}

\begin{figure}
    \centering
    \includegraphics[width = 0.4\textwidth]{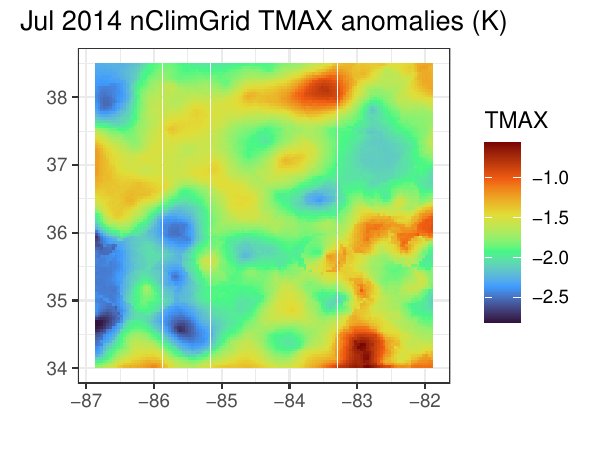}
    \includegraphics[width = 0.4\textwidth]{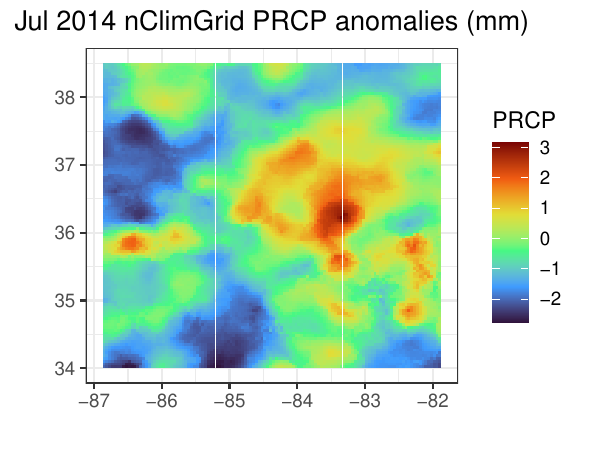}    
    \includegraphics[width = 0.4\textwidth]{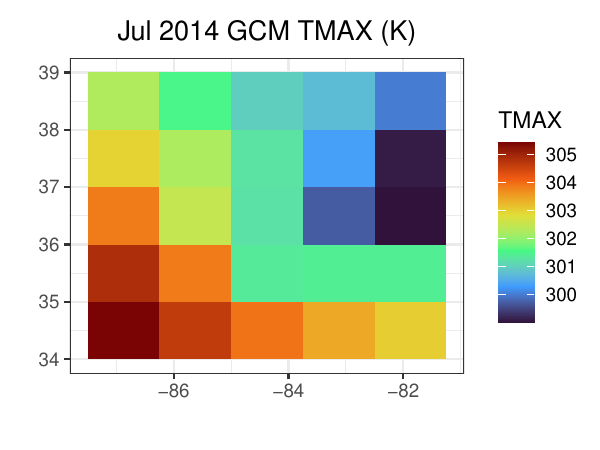}
    \includegraphics[width = 0.4\textwidth]{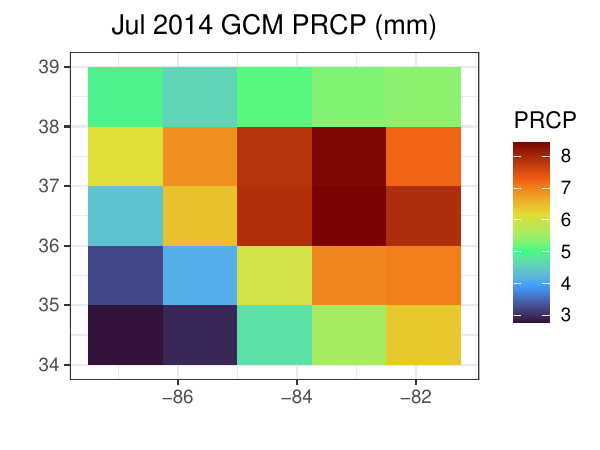}
       \caption{Monthly mean fields of nClimGrid anomalies (top) and GCM raw data (bottom) of TMAX and PRCP for the Southeast US study region.}
    \label{f:values}
\end{figure}

The study utilizes an ensemble member from the Geophysical Fluid Dynamics Laboratory (GFDL) GCM of of the CMIP6 experiments. The model, developed by NOAA, provides data at a spatial resolution of $100\,\mbox{km} \times 100 \,\mbox{km}$, which are used to analyze and density-correct the joint distribution of daily TMAX and PRCP for selected grids in the Southeast US and the Southwest US which have very distinct climatology. The Southwest grid (Figure \ref{f:locations}, left panel) in the four corners region is arid and has complex topography, spanning the states of Utah, Colorado, Arizona, and Mexico. The Southeast grid (Figure \ref{f:locations}, right panel) is humid, and spans the states of Indiana, Ohio, West Virginia, Kentucky, Virginia, Tennessee, North Carolina, South Carolina, Alabama, and Georgia. The GCM data are obtained from Earth System Grid Federation (ESGF) World Climate Research Programme (WCRP). The GFDL model is meant to be representative of CMIP6, and our methodology can be applied to any GCM in principle. The primary observational data comes from NOAA's nClimGrid-daily data product \citep{NClimGrid,Durre_2022}, which offers daily precipitation and temperature data from 1951 to the present at a $5\mbox{ km} \times 5 \mbox{ km}$ resolution. The observed climate data is aggregated to the same resolution as the GCM data, such that all the datasets used in our study are spatially aligned. The aggregation is carried out by taking the mean of the nClimGrid pixels using the Climate Data Operators software \citep[CDO;][]{schulzweida_2023_10020800}. Figure \ref{f:values} plots the monthly mean fields for July 2014 TMAX and PRCP in the Southeast region. The top row shows nClimGrid anomalies, calculated by subtracting the long-term mean of the data, at its native resolution. The bottom row shows the raw GCM data for the same time period. We use 1951--2014 as our study period, since the CMIP6 historical simulations run until 2014, and the observational record begins at 1950.

\subsection{Density correction model for GCM projections at a single location}\label{s:mv}
Our proposed density correction model has three steps - estimation, projection, and calibration. We begin by stating our model for the general multivariate case in this section; explicit formulations for spatial and spatiotemporal data are described subsequently. 
We build a statistical model for $p$ continuous responses, $\bY=(Y_1,...,Y_p)$ at a single location, which are assumed to have invertible marginal quantile and distribution functions that depends on variables $\bZ$ (e.g., season or elevation). The model is described simultaneously for both climate model output and observational data, with an auxiliary variable $X=0$ indicating the observation is from the climate model and $X=1$ indicating it is from the observational record.

We write the joint distribution of $\bY$ as a product of univariate conditional distributions
%, which are themselves approximated by a Vecchia approximation \citep{vecchia1988estimation,stein2004approximating}:
\begin{equation}\label{e:fulllik}
    f(\bY|\bZ,X) = 
    %\prod_{j=1}^p f_j(Y_j|Y_{j-1},\ldots,Y_1,\bZ,X) \approx 
    \prod_{j=1}^p f_j(Y_j|\bY_{(j)},\bZ,X),
\end{equation}
where $\bY_{(j)} = \{Y_i:i\in \mathcal{N}_j'\}$ and the conditioning set is $\mathcal{N}_j' = \{1,...,j-1\}$, which depends on the ordering of the $p$ variables. 
%we refer the reader to \citet{guinness2018_kl} and \citet{katzfuss2021general} for discussions on optimal choices for cardinality and members of $\mathcal{N}_j$. The Vecchia approximation facilitates the decomposition of a complex, high dimensional response vector into a product of independent univariate likelihoods; we refer to $\bY_{(j)}$ as the \textit{Vecchia neighbor set} going forward. It is important to note that the approximation is applied only to the dependence structure within $\bY$, and the right hand expression of \eqref{e:fulllik} is still a valid joint distribution for $\bY$.
The responses $\bY$ can alternatively be represented as transformations of the latent variables $\bU = (U_1,...,U_p)$, with $U_j\iid\mbox{Uniform}(0,1)$ for $j\in\{1,...,p\}$.  The map from $\bU$ to $\bY$ can be expressed sequentially,
\begin{equation}\label{e:U2Y}
    Y_j = Q_j(U_j;\bU_{(j)},\bZ,X),
\end{equation}
for $j\in\{1,...,p\}$, with $\bU_{(j)}$ defined analogous to $\bY_{(j)}$. In \eqref{e:U2Y}, $Q_j$ is the univariate conditional quantile function of $Y_j$ given $\bY_{(j)}$, $\bZ$, and $X$, and thus encapsulates the relationships between the $p$ variables and their dependence on $X$.  Collectively, the transformation is written $\bY = Q(\bU;\bZ,X)$ with the element $j$ of $\bY$ given by \eqref{e:U2Y}.

The inverse map from $\bY$ to $\bU$ is determined by the cumulative distribution functions, $F_j$, corresponding to the $Q_j$; i.e., $F_j$ is the inverse function of $Q_j$, $F_j = Q_j^{-1}$.  The inverse map for response $j$ is
\begin{equation}\label{e:Y2U}
    U_j = F_j(Y_j;\bY_{(j)},\bZ,X).
\end{equation}
Collectively, the transformation is written as $\bU = F(\bY;\bZ,X)$, with the element $j$ of $\bU$ determined by \eqref{e:Y2U}. 

Given the functions $F$ and $Q$, calibration of climate model output is straightforward. Say $\bY$ is a realization from the climate model, i.e., $X=0$. Then $\bU$ can be computed from \eqref{e:Y2U} as $\bU = F(\bY;\bZ,0)$, and the calibrated value for the model variables is then $\bY^* = Q(\bU;\bZ,1) = (Y_1^*,...,Y_{p}^*)$. Combining these two steps gives the calibration function $C$:
\begin{equation}\label{e:Y2Ystar}
  \bY^* =  C(\bY;\bZ) := Q\{F(\bY;\bZ,0);\bZ,1\}.
\end{equation}
If $F$ and $Q$ are correct, then the calibrated climate model output has the same joint distribution as the station data.

\begin{algorithm}[t]
\caption{Density correction of $Y_1=$ TMAX and $Y_2=$ PRCP at a single location}
\label{a:alg1}
\begin{algorithmic}
    \Require Data $\bY, \bZ,X$
    \State $\bY_{(2)} \gets Y_1$
    \Procedure{SPQR for TMAX and PRCP}{}
    \State Estimate $F_1$ and $Q_1$ for $Y_1|\bZ,X$ \Comment{Estimation step}
    \State Estimate $F_2$ and $Q_2$ for $Y_2|\bY_{(2)},\bZ,X$ \Comment{Estimation step}
    \EndProcedure
    \Procedure{TMAX density correction}{} 
        \State $u_1 \gets F_1(Y_1|\bZ, X=0)$ \Comment{Projection step}
    \State $Y_1^* \gets Q_1(u_1|\bZ, X=1)$ \Comment{Calibration step}
    \EndProcedure
    \Procedure{PRCP density correction}{}
    \State $u_2 \gets F_2(Y_2|\bY_{(2)},\bZ, X=0)$ \Comment{Projection step}
    \State $\bY_{(2)}^* \gets Y_1^*$
    \State $Y_2^* \gets Q_2(u_2|\bY_{(2)}^*, \bZ,X=1)$ \Comment{Calibration step}
    \EndProcedure
\end{algorithmic}
\end{algorithm}

While (\ref{e:Y2Ystar}) holds for any ordering of the $p$ variables, certain orderings will likely to lead to better estimation depending on the complexity of the $p$ densities. In the simple case of TMAX and PRCP at a single location (i.e., $p=2$), a natural ordering would be based on choosing $Y_1$ as TMAX and $Y_2$ as PRCP. This decomposes the conditional distribution of PRCP and TMAX as a conditional and a marginal distribution, assuming PRCP depends on TMAX. Physically, TMAX and PRCP are coupled, and choosing $Y_1$ as PRCP and $Y_2$ as TMAX also provides a valid statistical formulation. However, we assume that PRCP depends on TMAX for somewhat practical reasons. Intuitively, the distribution of TMAX has a much simpler functional form compared to PRCP. PRCP has a preponderance of zeros, and is therefore harder to estimate compared to TMAX. Our proposed dependence structure mitigates this by using TMAX as a predictor for PRCP. Our formulation is also validated in practice; SPECD where TMAX depended on PRCP led to noticeably worse predictions over both the training and validation periods.

The sequential formulation reduces calibration to estimating the univariate CDFs, $F_1,...,F_p$, sequentially. The function $F_j$ is the conditional CDF of $Y_j$ treating $\bY_{(j)}$, $\bZ$ and $X$ as features. These functions can be estimated in parallel with distributional regression methods that allow for a potentially large number of features and flexible relationship between the features and the response distribution. We use semi-parametric quantile regression (SPQR) for the distributional regression which is detailed in Section \ref{s:spqr}. Algorithm \ref{a:alg1} summarizes the calibration process for TMAX and PRCP at a single location.

\subsection{Extension to spatial data}\label{s:mvs}

\begin{algorithm}[t]
\caption{Density correction for $Y_1 = $ TMAX and $Y_2 = $ PRCP at multiple locations}
\label{a:alg2}
\begin{algorithmic}
    \Require Data $\bY,X$
    \For{$l = 1,\ldots,n$}
    \State $\bY_{(1l)} = \{Y_{1,l-1},\ldots,Y_{1,l-m}\}$
    \State $\bY_{(2l)} = \{Y_{1,l},\ldots,Y_{1,l-m},Y_{2,l-1},\ldots,Y_{2,l-m}\}$
    \Procedure{SPQR for TMAX and PRCP}{}
    \State Estimate $F_{1l}$ and $Q_{1l}$ for $Y_{1l}|\bY_{(1l)},X$ \Comment{Estimation step}
    \State Estimate $F_{2l}$ and $Q_{2l}$ for $Y_{2l}|\bY_{(2l)},X$ \Comment{Estimation step}
    \EndProcedure
    
    \Procedure{TMAX density correction}{} 
    \State $u_{1l} \gets F_{1l}(Y_{1l}|\bY_{(1l)},X=0)$ \Comment{Projection step}
    \State $\bY_{(1l)}^* = \{Y_{1,l-1}^*,\ldots,Y_{1,l-m}^*\}$
    \State $Y_{1l}^* \gets Q_{1l}(u_{1l}|\bY_{(1l)}^*,X=1)$ \Comment{Calibration step}
    \EndProcedure
    \Procedure{PRCP density correction}{}
    \State $u_{2l} \gets F_{2l}(Y_{2l}|\bY_{(2l)},X=0)$ \Comment{Projection step}
    \State $\bY_{(2l)}^* = \{Y_{1,l}^*,\ldots,Y_{1,l-m}^*,Y_{2,l-1}^*,\ldots,Y_{2,l-m}^*\}$
    \State $Y_{2l}^* \gets Q_{2l}(u_{2l}|\bY_{(2l)}^*,X=1)$ \Comment{Calibration step}
    \EndProcedure
     \EndFor
\end{algorithmic}
\end{algorithm}

Consider our problem of interest - density correction of TMAX and PRCP at multiple locations. If there are $n$ locations, a reasonable approach would be to first order them (e.g., based on their geographical coordinates), and then order TMAX and PRCP at each location. This could be represented by $\bY = (Y_1,\ldots,Y_p)$ for $p=2n$, and theoretically, Algorithm \ref{a:alg1} from Section \ref{s:mv} can be applied. However, as the number of locations increase, so does the number of features $\bY_{(j)}$. This soon leads to computational problems, and directly applying \eqref{e:fulllik} to large spatial (and spatiotemporal) data is computationally prohibitive. To mitigate this, we approximate \eqref{e:fulllik} by means of a Vecchia approximation \citep{vecchia1988estimation,stein2004approximating,datta2016a,katzfuss2021general}:
\begin{equation}\label{e:vecchia}
    f(\bY|\bZ,X) = 
    \prod_{j=1}^p f_j(Y_j|Y_{j-1},\ldots,Y_1,\bZ,X) \approx 
    \prod_{j=1}^p f_j(Y_j|\bY_{(j)},\bZ,X),
\end{equation}
where $\bY_{(j)} = \{Y_i;i\in \mathcal{N}_j\}$ and the conditioning set is $\mathcal{N}_j \subseteq \mathcal{N}_j' = \{1,...,j-1\}$. We refer to $\bY_{(j)}$ as the \textit{Vecchia neighbor set} going forward. The size of the conditioning set $\mathcal{N}_j$ is therefore bounded above, i.e., $|\mathcal{N}_j|\leq j-1$. In practice, $10 \leq |\mathcal{N}_j|\leq 30$ are often sufficient even when $p$ is very large; we refer the reader to \citet{guinness2018_kl} and \citet{katzfuss2021general} for discussions on optimal choices for cardinality and members of $\mathcal{N}_j$. The Vecchia approximation facilitates the decomposition of a complex, high-dimensional response vector into a product of independent univariate likelihoods. It is important to note that the approximation is applied only to the dependence structure within $\bY$, and the right hand expression of \eqref{e:vecchia} is still a valid joint distribution for $\bY$.

To apply the Vecchia decomposition to the joint distribution of TMAX and PRCP at $n$ locations (with the extension to more than two variables being straightforward), let $Y_{1l}$ and $Y_{2l}$ denote the response for TMAX and PRCP at location $l \in \{1,\ldots,n\}$ corresponding to grid point $\bs_l = (s_{l1},s_{l2})$. The $p=2n$ variables are therefore ordered as $\bY = (Y_{11},Y_{21},\ldots,Y_{1n},Y_{2n})$. 
In this study, we choose the Vecchia neighboring set for $Y_{1l}$ to be $\bY_{{(1l)}} = \{Y_{1,l-1},\ldots,Y_{1,l-m}\}$, and the Vecchia neighboring set for $Y_{2l}$ as $\bY_{{(2l)}} = \{Y_{1l},\ldots,Y_{1,l-m},Y_{2,l-1},\ldots,Y_{2,l-m}\}$. This encapsulates our prior assumption that PRCP depends on TMAX but not vice versa. Therefore, while TMAX at a location depends on TMAX at its Vecchia neighbor locations, PRCP depends on TMAX at the current location and the PRCP at the (same) Vecchia neighbor locations. While other formulations are possible, we found this to be intuitive and tractable. Algorithm \ref{a:alg2} summarizes the calibration process for TMAX and PRCP at multiple locations.

Note that it is possible to extend the ideas of the spatial setting into a spatiotemporal version of the algorithm, where time-lagged variables are used as covariates to capture autocorrelation. Appendix \ref{s:mvst} outlines this extension to the spatiotemporal setting.

\subsection{Conditional density estimation using SPQR}\label{s:spqr}
Instead of prescribing parametric forms to the univariate conditional distributions $f_j$ in \eqref{e:fulllik} and \eqref{e:vecchia}, we use semi-parametric quantile regression \citep[SPQR,][]{xu-reich-2021-biometrics} to obtain $F$, $Q$, and subsequently, $C$. For simplicity, in the remainder of this section, we use $Y$ and $\bX$ respectively to denote $Y_j$ and $(\bY_{(j)}, X, \bZ)$ as in \eqref{e:fulllik} and \eqref{e:vecchia}.
SPQR assumes the PDF of $Y|\bX$ to be a function of $K$ second-order M-spline basis functions, $B_1(Y),...,B_K(Y)$, with equally-spaced knots spanning $[0,1]$. Each M-spline basis function is a valid PDF on $[0,1]$ \citep{ramsay1988}, and therefore so are their convex combinations. The SPQR model is 
\begin{equation}\label{e:spqrlik}
 f(Y|\bX,\phi) = \sum_{k=1}^K\pi_k(\bX,\phi)B_k(Y),
\end{equation}
where the probabilities $\pi_k(\bX,\phi)$ function as weights that satisfy $\pi_k(\bX,\phi)\ge 0$ and $\sum_{k=1}^K\pi_k(\bX,\phi)=1$, and $\phi$ are trainable neural network parameters defined below.

By increasing the number of basis functions $K$ and appropriately selecting the weights, the mixture distribution in \eqref{e:spqrlik} can approximate any continuous density function \citep[e.g.,][]{chui1980,abrahamowicz1992}. To ensure that the weights are chosen in a flexible manner, they are modeled using a multi-layer perceptron (MLP) neural network. The MLP has H hidden layers with rectified linear unit \citep[ReLU,][]{NairHinton2010} \textit{activation functions}, which aim to represent nonlinear relationships between the inputs and outputs of each layer. The final layer has a softmax (or equivalently, an expit) activation function, which ensures that the output is a vector of probabilities that sum to 1. The output of layer $h$ is:
\begin{align}
    \bx^{(h)} &= \mbox{ReLU}(\bW^{(h)}\bx^{(h-1)} + \bb^{(h)}),\,h=1,\ldots,H,\\
    \Pi_K(\bX) := \bx^{(H+1)} &= \mbox{softmax}(\bW^{(H+1)}\bx^{(H)} + \bb^{(H+1)}),
\end{align}
where $\bx^{(0)}:=\bX$ corresponds to the input covariates, and $\Pi_K(\bX) = \{\pi_k(\bX,\phi)\}_{k=1}^K$ is a $K-$vector of probabilities that sum to 1. The estimable parameters of the MLP are $\phi = \{\{\bW^{(h)},\bb^{(h)}\}_{h=1}^{H+1}\}$, where $\bW^{(h)} \in \mathbb{R}^{n_h \times n_{h-1}}$ and $\bb^{(h)} \in \mathbb{R}^{n_h}$ are layer-specific weight matrices and bias vectors, respectively. The dimension $n_h$ is referred to as the number of \textit{nodes} in a layer. If we represent each layer operation by a function $g^{(h)}(\cdot)$, suppressing the dependence on $\phi$ and $\bX$ for convenience, the MLP $g:\mathbb{R}^q \rightarrow \mathbb{R}^K$ is a composition of the layer operations, i.e., $g(\cdot) := g^{(H+1)} \circ \ldots \circ g^{(1)}(\cdot)$.

The weights and biases in $\phi$ are optimized through use of stochastic gradient descent with the adaptive moment estimation (Adam) optimizer \citep{kingma2014adam}; the negative log-likelihood associated with \eqref{e:spqrlik} serves as the loss function for the optimization.
Computing the PDF is simple and fast given the parameters $\phi$. Additionally, the integral of M-spline functions are I-spline functions \citep{ramsay1988}, a fact that is exploited to obtain an expression for the cumulative distribution function (CDF):
\begin{equation}\label{e:CDF}
 F(Y|\phi,\bX) = \sum_{k=1}^K\pi_k(\bX,\phi) I_k(Y),
\end{equation}
where $I_k(Y)$ are I-spline basis functions. The conditional quantile function for quantile level $\tau\in(0,1)$ is defined as the function $Q(\tau|\phi,\bX)$  so that 
$$F\{Q(\tau|\phi,\bX)|\phi,\bX\}=\tau.$$ The conditional quantile function for this model is not available in closed-form, but can be approximated by numerically inverting $F(Y|\phi,\bX)$. Given that \eqref{e:CDF} models a valid CDF, the conditional quantile function estimated through this approach satisfies the non-crossing constraint $$\frac{\partial Q(\tau|\phi,\bX)}{\partial\tau}>0,\ \forall\ \bX,$$ and does not require any second-stage monotonization treatment \citep{bondell2010noncrossing}.

As previously mentioned, SPQR is a method to approximate continuous density functions. However, daily PRCP contains a large proportion of zeroes making it a semi-continuous variable. Therefore, instead of modeling PRCP directly, we model $\log(0.0001 + \mbox{PRCP})$, a common approach when modeling precipitation data using continuous distributions \citep{Rajagopalan1998}. We found this transformation adequate for estimating the proportion of zeros in the data, combined with a post-processing step (described in Section \ref{s:app:results}). However, SPQR can underestimate high quantiles of heavy-tailed distributions, since the \textit{M}-splines are bounded and therefore not amenable to extrapolation. We noted this issue in our numerical experiments and in the discussion section.

\subsection{Model evaluation metrics for SPECD}\label{s:gof}
In the studiers presented in Sections \ref{s:sim}--\ref{s:app}, we assess the efficacy of SPECD based on three functionals each of the marginal distributions and the dependence structure of calibrated PRCP and TMAX. We evaluate the Wasserstein distance \citep{kantorovich,wasserstein} between the marginal densities of the calibrated and observational data, which quantifies the goodness-of-fit for each variable. Similarly, we also evaluate the proportion of zeros for PRCP and the 0.95 quantile for PRCP and TMAX. These two quantities focus on the tail behavior of the two variables. For assessing whether SPECD can preserve the dependence structure within nClimGrid, we evaluate pairwise spatial correlations for both variables, and the cross correlation between PRCP and TMAX. For all metrics other than the Wasserstein distance, we report the mean absolute error (MAE) between the observed and the calibrated data.

In addition to assessing goodness of fit of our density corrected data, we also compare it with QM and CCA. As described in Section \ref{s:intro}, 
QM is a bias correction method which matches the quantiles of the observational data time series and the GCM time series. The asynchronous QM bias correction method sorts GCM and nClimGrid data vectors in ascending order (independently), and then fits linear regression to the sorted data. This can take the form of simple linear regression \citep[e.g.,][]{Dettinger2004} or piecewise linear regression \citep{stoner2013}. The QM methodology is applied separately for each month, grid cell, and variable (TMAX and PRCP). The goal is to obtain a linear map between the quantiles of the GCM and nClimGrid datasets. We can then provide GCM data for unobserved grid points or time periods as input to the fitted model and obtain bias-corrected TMAX and PRCP as output. 
Since QM in our example is applied to each variable separately, it fails to account for inter-variable dependencies \citep{Bhowmick2019}. This can distort relationships critical for understanding processes such as hydrological cycles \citep{SEO2019304}. We use CCA \citep{bhowmick2017} as our other point of comparison. CCA employs multivariate sorting of variables based on their joint probability of occurrence such that their cross correlation is maintained. Afterwards, canonical correlation is used to bias correct variables simultaneously. Both CCA and QM are applied independently to each grid cell, and therefore do not account for spatial correlation in the process of bias correction. Similarly, the asynchronous nature of both algorithms lead to temporal dependence is not specifically accounted for. Both methods have to contend with a well known phenomenon in GCMs called the \textit{drizzle effect} \citep[see, e.g.,][]{levey2024}, where numerical climate models for precipitation very rarely generate zero precipitation values, instead usually having very small but positive values. QM and CCA address this by applying a wet-day fraction correction in the form of a random offset added to the zero precipitation values in the observational data \citep{Pierce2015,Wootten2021}. The offset for each zero precipitation value is drawn independently from a Uniform$(0.001,0.1)$ distribution, leading to unique non-zero precipitation values. After bias corrections, precipitation values smaller than $0.01$ mm is set to zero.

\section{Numerical Study}\label{s:sim}
\subsection{Overview}
We study the efficacy of SPECD using a numerical study whose data generating process approximates our application. Since we will be fitting SPECD for each month separately, the numerical study focuses on modeling a single 30-day month's TMAX and PRCP data, available for a period of 64 years. The first 50 years are treated as training data and is used to fit the model. Density correction is carried out on all 64 years of data, with the first 50 years used to benchmark in-sample performance while the last 14 years used to measure out-of-sample performance. Calibrating the data that the model is trained on is analogous to density correction of historical data. In contrast, out-of-sample performance gauges the method's ability for calibrating future projections, provided the distribution of the GCM data does not change between the two time periods.\\

\begin{figure}
    \centering
    \includegraphics[scale=0.4]{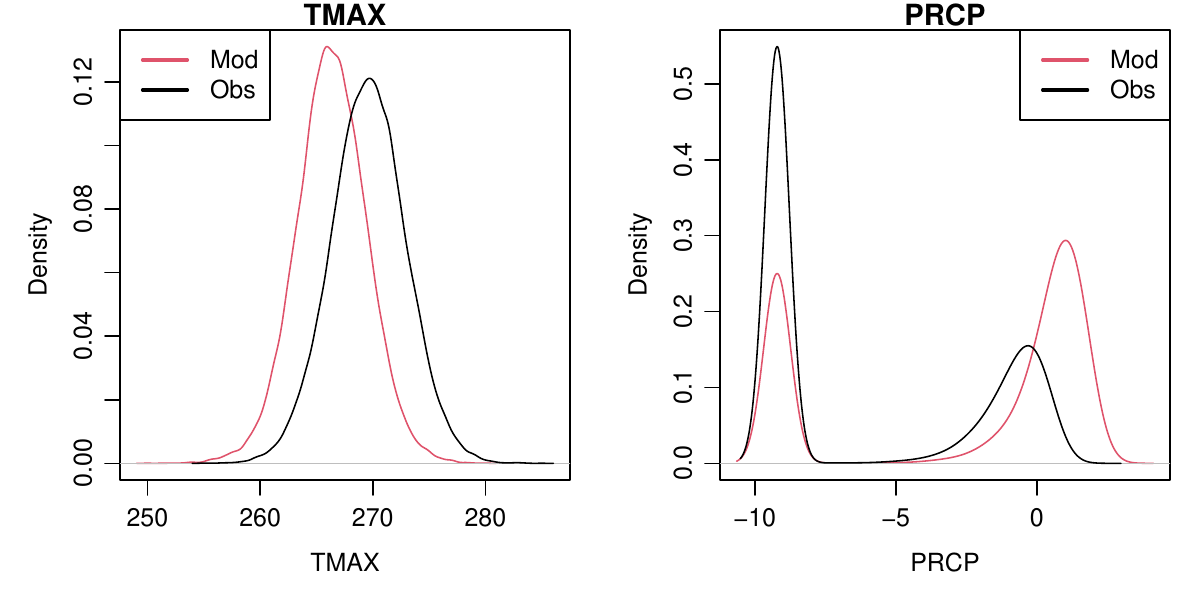}
    \caption{Density curves of climate model (Mod) and observed (Obs) TMAX and PRCP data, based on a single replicate from the simulation study. PRCP curve is presented on the log scale.}
    \label{f:density_sim}
\end{figure}

\noindent \textbf{Data generating process:} We generate the data for $64\times 30$ days at $m = 25$ locations $\{1,2,3,4,5\}^2$, i.e., on a $5\times 5$ grid. For spatial location $i$ and replicate $t$, there are four responses, meant to represent (1) observed temperature, (2) observed precipitation, (3) model temperature and (4) model precipitation.   The $4m$ outcomes for each replicate are generated from multivariate skew-t distribution. The location parameter is $(1/m,...,m/m) \otimes (2,3,1,2)$ so that each outcome has a different spatial trend.  The scale matrix is $\Omega\otimes R$ where $\Omega$ is the spatial correlation matrix whose $(i,j)^{th}$ element is equal to $\exp(-|i-j|/2)$ and $R$ is the $4\times4$ correlation matrix with elements $R_{12} = -0.8$, $R_{34}=-0.4$ and all other off-diagonal elements have a magnitude of 0.5. The strength of dependence between temperature and precipitation is stronger in the observations than the model under this setup. The degrees of freedom is set to 20; the skewness is 0 for both temperature responses, 100 for observed precipitation, and 10 for model precipitation. This results in non-Gaussian data with more skewness in the observations than the model. After generating the replications from the multivariate skewed-t distribution, we apply a spatial smoother (Gaussian kernel with bandwidth 2) to variables representing model output so that the spatial dependence is stronger in model output variables. 

Finally, we rescale both TMAX and PRCP to ensure their scales are similar to what we observe in the application. The observational TMAX is scaled using a min-max transform to be between $255-285\, K$, while the model TMAX is scaled to be between $250-280\, K$. For observational PRCP, we shift the center of the distribution to the $75^{th}$ quantile, and threshold everything below that to be $0$. A similar transformation is carried out for the model PRCP, but using the median value. As a consequence, $75\%$ of the observational PRCP is $0$, and half of the model PRCP is $0$, reflecting real world scenarios where GCMs have far fewer zeros in the data compared to the observational record.
Figure \ref{f:density_sim} plots the density curve for the 64 years of data based on one simulated dataset, pooled over the 25 grid cells. We see that the generated process emulates a hotter and drier month in the observational record compared to the climate model output.

\noindent \textbf{Models and testing criteria:} We compare four different bias/density correction approaches based on the metrics laid out in Section \ref{s:gof}. The first two are classical QM and CCA. The third and fourth are variants of SPECD, corresponding to Algorithms \ref{a:alg1} and \ref{a:alg2}, respectively. SPECD1, described in Section \ref{s:mv}, models all locations independent of each other, but considers the cross correlation between TMAX and PRCP. SPECD2 is described in Section \ref{s:mvs} and considers both spatial correlation within each variable and cross correlation between the variables. The two variants of SPECD aim to disambiguate potential improvements in calibration due to SPQR from those due to the spatial information encoded via the Vecchia approximation.

\subsection{Results}

\begin{figure}
     \centering
     \begin{subfigure}[b]{\textwidth}
         \centering
        \includegraphics[scale=0.5]{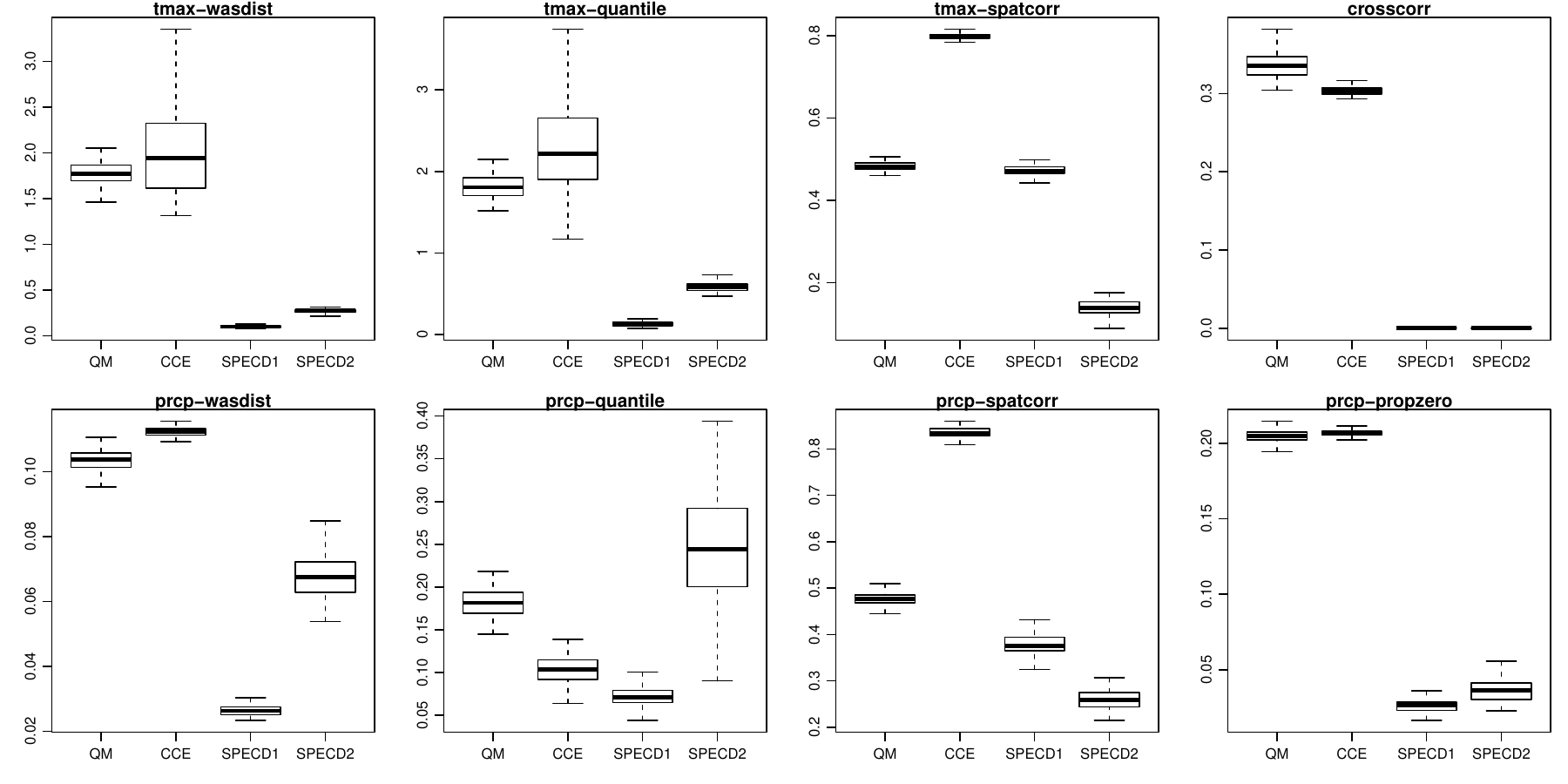}
         \caption{Simulation study results for training period of 1951--2000.}
         \label{fig:sim_results_train}
     \end{subfigure}
     \hfill
     \begin{subfigure}[b]{\textwidth}
         \centering
         \includegraphics[scale=0.5]{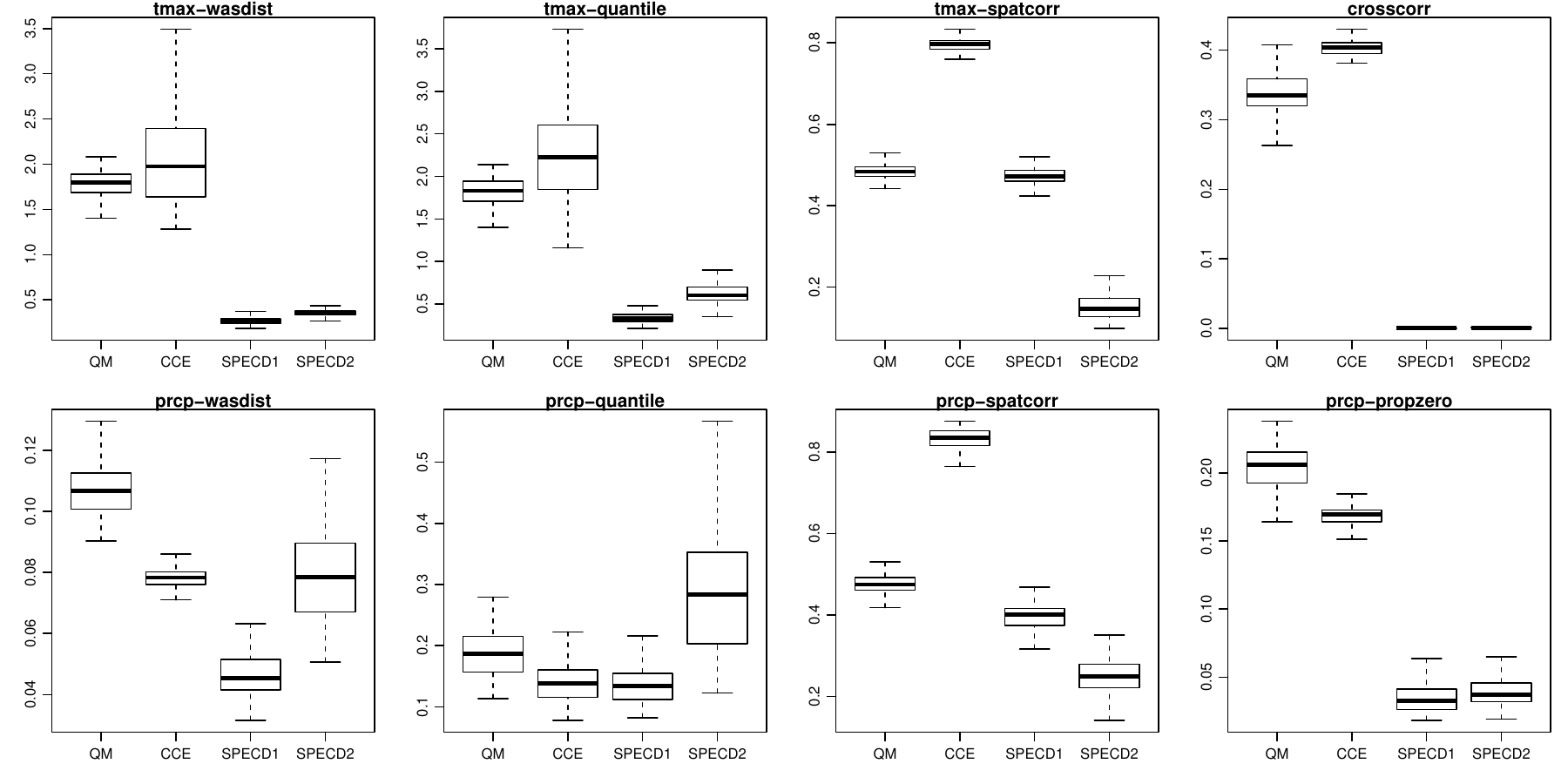}
         \caption{Simulation study results for validation period of 2001--2014.}
         \label{fig:sim_results_test}
     \end{subfigure}
        \caption{Comparison of QM and CCE against two SPECD methods without (SPECD1) and with (SPECD2) spatial correlation. The left column reports Wasserstein distance; the remaining columns report MAE. Lower values indicate greater accuracy.}
        \label{fig:sim_results}
\end{figure}

Figure \ref{fig:sim_results} contains boxplots of the metrics detailed in Section \ref{s:gof} based on the 100 datasets. The top rows in Figures \ref{fig:sim_results_train} and \ref{fig:sim_results_test} each contain three metrics for TMAX - the Wasserstein distance between the true and calibrated densities (\texttt{tmax-wasdist}), and MAEs for the $95^{th}$ quantile (\texttt{tmax-quantile}) and the spatial correlation between the grid cells (\texttt{tmax-spatcorr}). The final plot reports the MAE of the cross correlations between the variables (\texttt{crosscorr}). The bottom rows each contain four panels of PRCP metrics - the Wasserstein distance between the true and calibrated densities (\texttt{prcp-wasdist}), and MAEs for the $95^{th}$ quantile (\texttt{prcp-quantile}), the spatial correlation (\texttt{prcp-spatcorr}), and the proportion of zeros (\texttt{prcp-propzero}). Each individual metric is evaluated based on the entire data - i.e., 50 years for Figure \ref{fig:sim_results_train} and 14 years for Figure \ref{fig:sim_results_test}. Additionally, for all metrics except for the spatial correlation, the metrics are also pooled over the 25 locations.

We begin by noting that Figure \ref{fig:sim_results_train} has smaller values overall than Figure \ref{fig:sim_results_test}, confirming that in-sample performance is better than out-of-sample performance. There is also increased variability in Figure \ref{fig:sim_results_test}, reflecting higher uncertainty in out-of-sample estimates. However, the two figures demonstrate broadly similar patterns otherwise. For the three TMAX metrics, both SPECD approaches outperform QM and CCA by significant margins. SPECD2 has the lowest MAE for spatial correlation in TMAX compared to the other methods; however, SPECD1 has slightly lower Wasserstein distance and in upper quantile MAE. Both SPECD methods have significantly lower cross correlations compared to QM and CCA.

For PRCP, SPECD still outperforms QM and CCA overall. However, SPECD2 has the highest MAE for the upper quantile than any of the competing methods, whereas SPECD1 has the lowest. We believe this is due to SPECD2 being the only method which explicitly models the spatial dependence structure. As a consequence, the marginal errors are somewhat higher. This is further exacerbated by PRCP being modeled on the log-scale, as errors get magnified when transformed back to the original scale. SPECD1, which only considers individual grid points, does not suffer from this issue. SPECD1 also has the lowest Wasserstein distance, but SPECD2 has the lowest MAE in spatial correlation estimates due to its underlying assumptions. Both SPECD algorithms perform comparably in their estimates of the proportion of zeros in PRCP data.

Overall, we see that SPECD comprehensively outperforms QM and CCE while calibrating the joint distribution of TMAX and PRCP over both the training and validation periods. The only exception lie with the heavy tails of PRCP, where SPECD runs into some issues when both the response and covariates are on the log scale. A table with the mean and standard errors for all the estimates, based on the 100 datasets, is provided in Appendix \ref{s:AppB_sim}. We also note minimal loss in predictive accuracy when going from SPECD1 to SPECD2 for marginal quantities. SPECD2 still outperforms QM and CCA, but not has the added benefit of incorporating spatial information in a theoretically justified manner. 

\section{Density correction of historical TMAX and PRCP}\label{s:app}
\subsection{Overview}
We implemented SPECD for daily observational and GCM data from 1951--2014 on two separate $5 \times 5$ grids as shown in Figure \ref{f:locations}. We separated the data into a training period (1951--2000) and a validation period (2001--2014), and fitted the SPECD model (corresponding to SPECD2 from the numerical study) to the training period data. The Vecchia neighbor sets are selected using max-min ordering, which can result in significant improvements over other coordinate based orderings \citep{guinness2018_kl}.  

We model each month's data separately to account for seasonality; for each month and at every location conditioned on its neighboring set of $m=10$ locations, we fitted SPQR models for TMAX and the transformed PRCP. Each SPQR model has several hyperparameters, which were chosen based on a grid search across different hyperparameter values (not presented). We choose each SPQR model to have 2 hidden layers with 30 and 20 nodes, and 20 output knots for both variables; the hidden layers have ReLU activation functions \citep{NairHinton2010}. The batch size and learning rate is set to 100 and 0.001 respectively, and the SPQR model is run for 300 iterations. In each model, $20\%$ of the training data is used as a validation set, with early stopping enforced if the validation set loss fails to decrease for more than 5 iterations.

The SPQR models (for each month, location, and variable) can be fit in parallel, since the Vecchia approximation decomposes the joint density of $\bY$ into independent univariate conditional densities. Predictions for every month can also be carried out independently, and therefore the projection/calibration step occurs in parallel. Within each month, however, predictions need to be made in the same sequence as the ordering of the Vecchia neighbor sets. Predictions for TMAX and PRCP were obtained at each location using Algorithm~\ref{a:alg2}. 

A key requirement for evaluating predictive models based on out-of-sample performance is for the training and validation data to arise from the same population distribution. In the context of density correction, the spatial dependence structure should therefore be identical in the training (1951--2000) and validation (2001--2014) periods. 
An exploratory study in Appendix \ref{s:App_stationarity} examines this behavior, by evaluating and plotting the spatial correlation for each pair of locations, during the training and validation periods. These are done separately for every month, in line with the density correction that is done separately for each month. Ideally, we want the differences in the spatial correlations between the two periods to be centered at zero, and have low variability. However, our study indicates that neither of these two properties hold for large sections of our data. Spatial correlations for PRCP show higher variability between the two periods than TMAX, possibly due to the presence of zeros in the data. The differences were often biased away from 0, indicating that the correlations have either strengthened/weakened over time. There were also situations where the model data and the observational data were biased in different directions; e.g., the spatial correlation in the model data decreases over time, while it strengthens in the observational data. Deviations were more frequent and more pronounced for pairs of locations with relatively low spatial correlations. While both regions demonstrated these patterns, it was especially prominent for the SW. We believe this to be at least in part due to the complex topography of the four-corners region. As a consequence, we anticipate that spatial correlations for the validation period might not be adequately calibrated based on the fitted models. SPECD is likely to be disproportionately affected compared to classical methods in such scenarios, since neural networks can struggle in generalizing outside the range of the observed data, or in the case of covariate drift.
However, these issues are largely absent when we only consider in-sample predictions, i.e., when evaluating the density correction of data from 1951--2000. In situations with out-of-sample non-stationarity, SPECD can still realistically model spatial (and temporal) structures in the data, balancing flexibility and scalability. Section \ref{s:discussion} discusses potential approaches to mitigate the effects of nonstationarity that we observe in the current study.

\subsection{Evaluating density-corrected output based on SPECD}\label{s:app:results}

\begin{figure}
     \centering
     \begin{subfigure}[b]{\textwidth}
         \centering
         \includegraphics[scale=0.35]{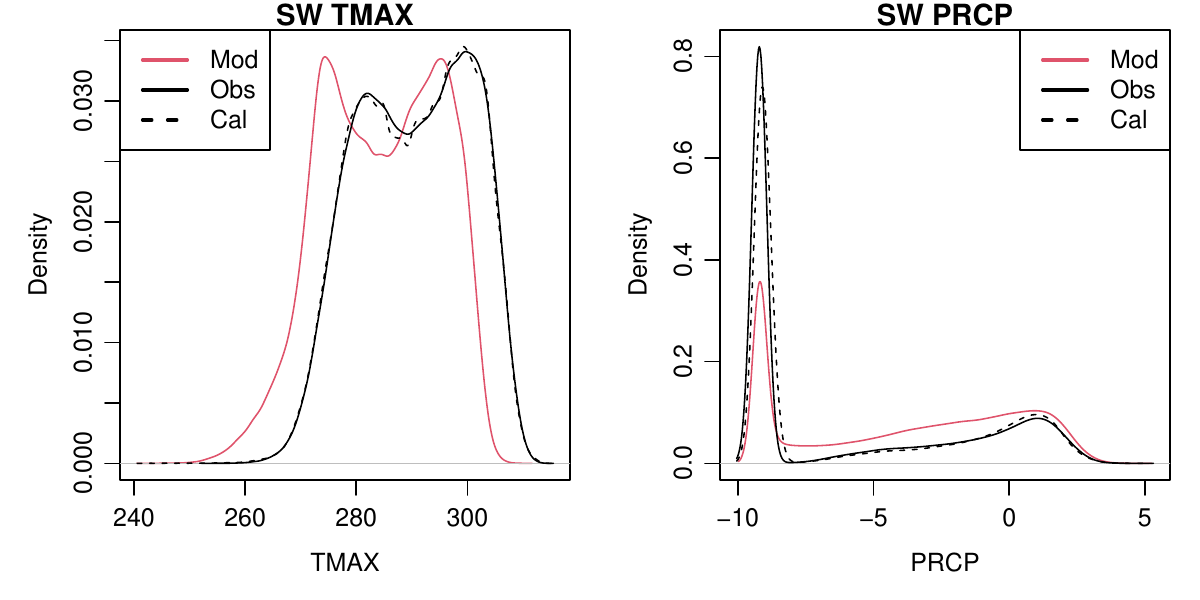}
    \includegraphics[scale=0.35]{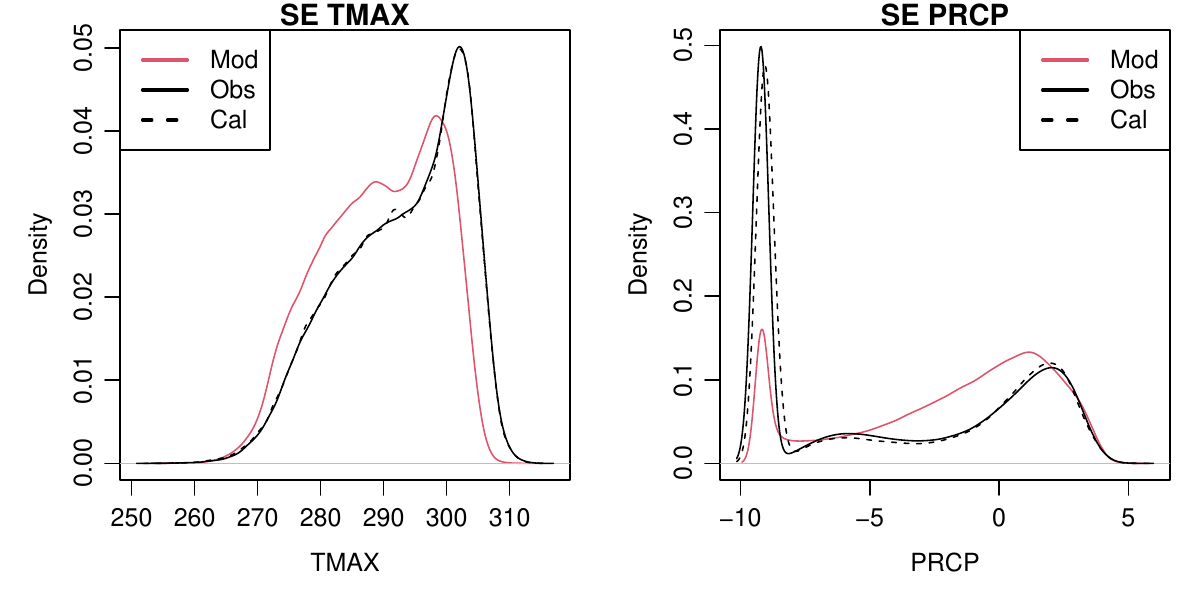}
         \caption{Density curves for training period of 1951--2000.}
         \label{fig:density_train}
     \end{subfigure}
     \hfill
     \begin{subfigure}[b]{\textwidth}
         \centering
         \includegraphics[scale=0.35]{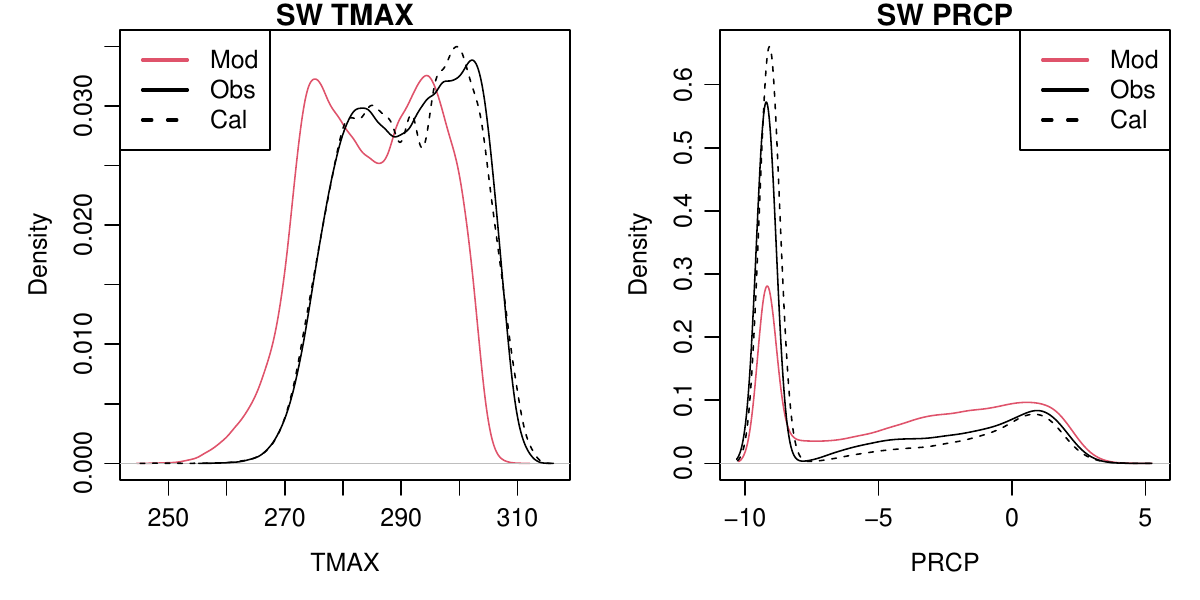}
    \includegraphics[scale=0.35]{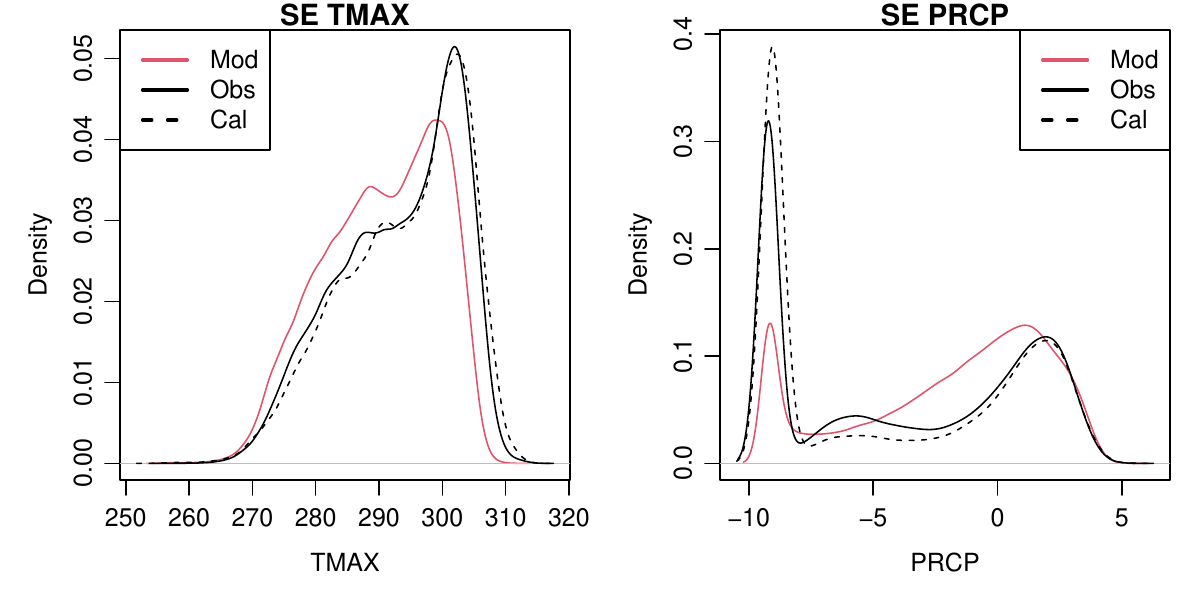}
         \caption{Density curves for validation period of 2001--2014.}
         \label{fig:density_validation}
     \end{subfigure}
        \caption{Marginal density curves of climate model (Mod), observed (Obs), and calibrated (Cal) TMAX and PRCP data in the Southwest (SW) and Southeast (SE), pooled across location and month. PRCP results are presented on the log scale.}
        \label{fig:density}
\end{figure}

In this section, we present results from density correction at both sets of locations for the training and validation periods. Figure \ref{fig:density} plots density curves of density corrected TMAX and log-scale PRCP for the SE and SW regions. Each curve pools the entire training/validation period data across all 25 locations, and we see that the calibrated TMAX and PRCP closely match the observed data for both time periods. Note that the density plots for PRCP are displayed in the transformed log scale since visual inspection of the density plots are most informative in the scale that the data was analyzed. However, all remaining plots and metrics for PRCP are presented in the raw data scale. Fitting separate monthly models for TMAX captures the pronounced bimodal distribution in the SW region with different modes for observed and model data. On the other hand, there are substantial differences between the observed and model densities of PRCP at both tails of their respective distributions. For both variables, SPQR is able to successfully calibrate the model data to match the density of the observational data.  

\begin{figure}
     \centering
     \begin{subfigure}[b]{\textwidth}
         \centering
         \includegraphics[scale=0.35]{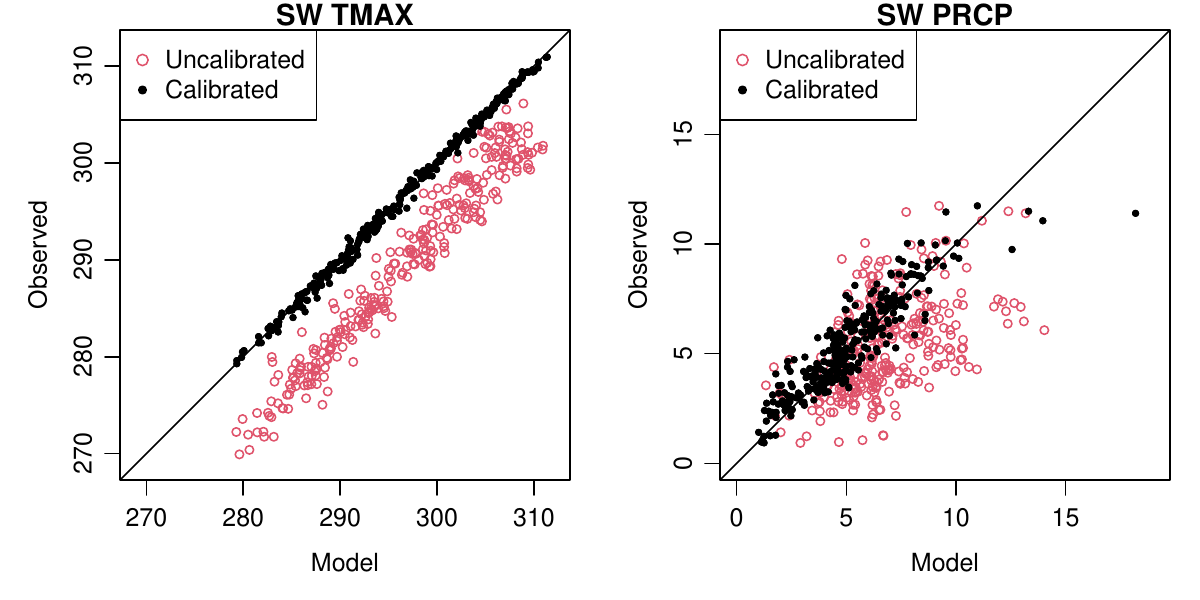}
        \includegraphics[scale=0.35]{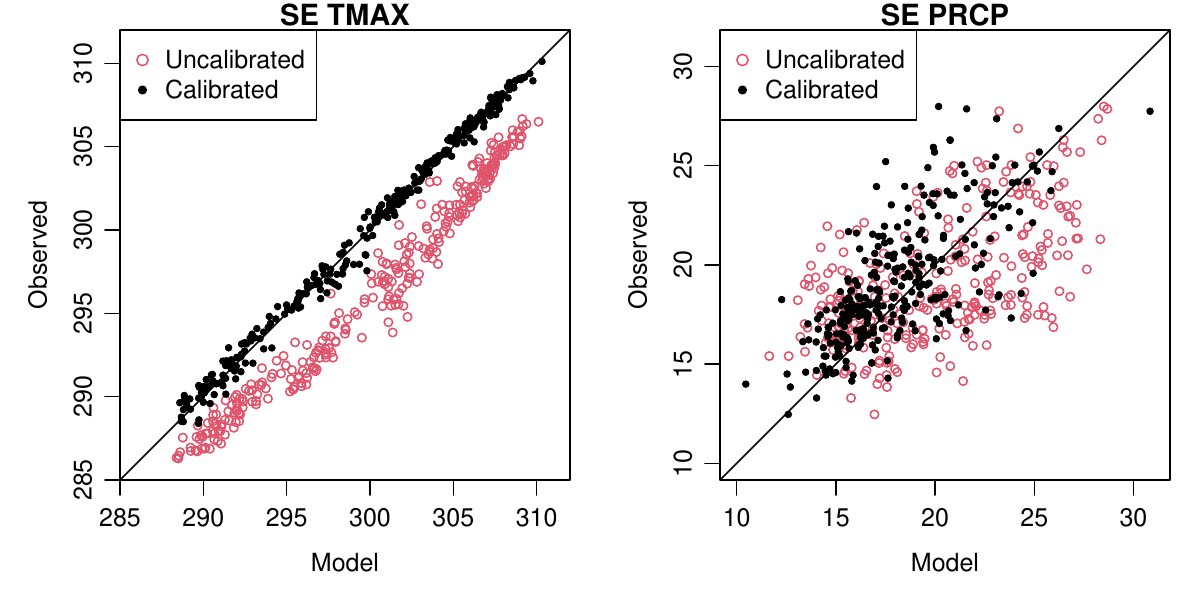}
         \caption{Monthly 0.95 quantiles for training period of 1951--2000.}
         \label{fig:summaries_train}
     \end{subfigure}
     \hfill
     \begin{subfigure}[b]{\textwidth}
         \centering
         \includegraphics[scale=0.35]{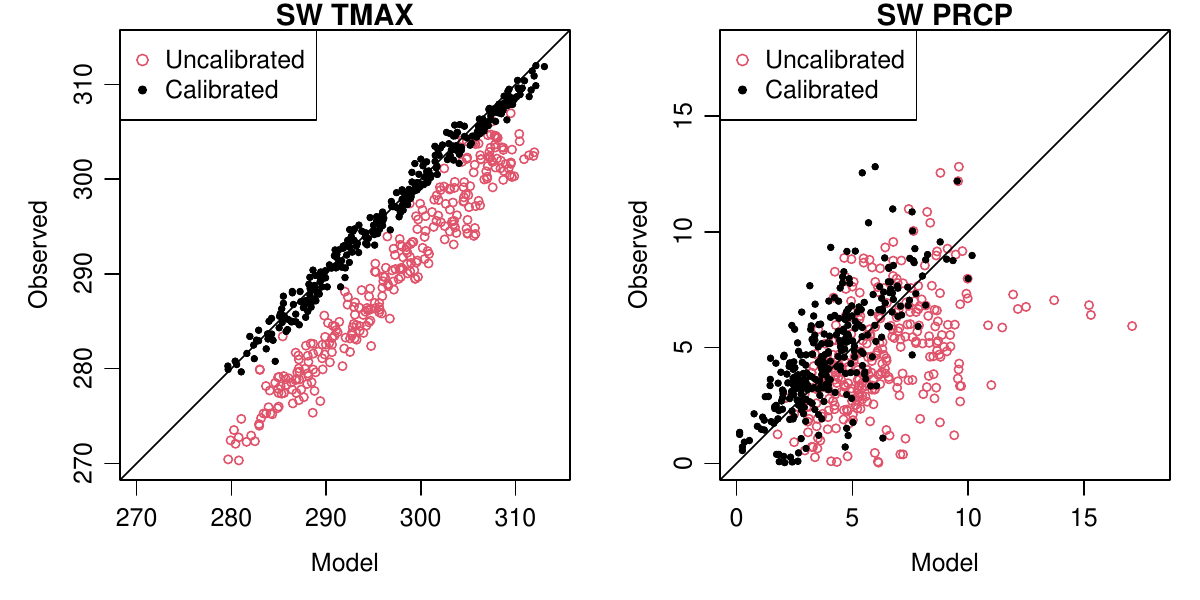}
        \includegraphics[scale=0.35]{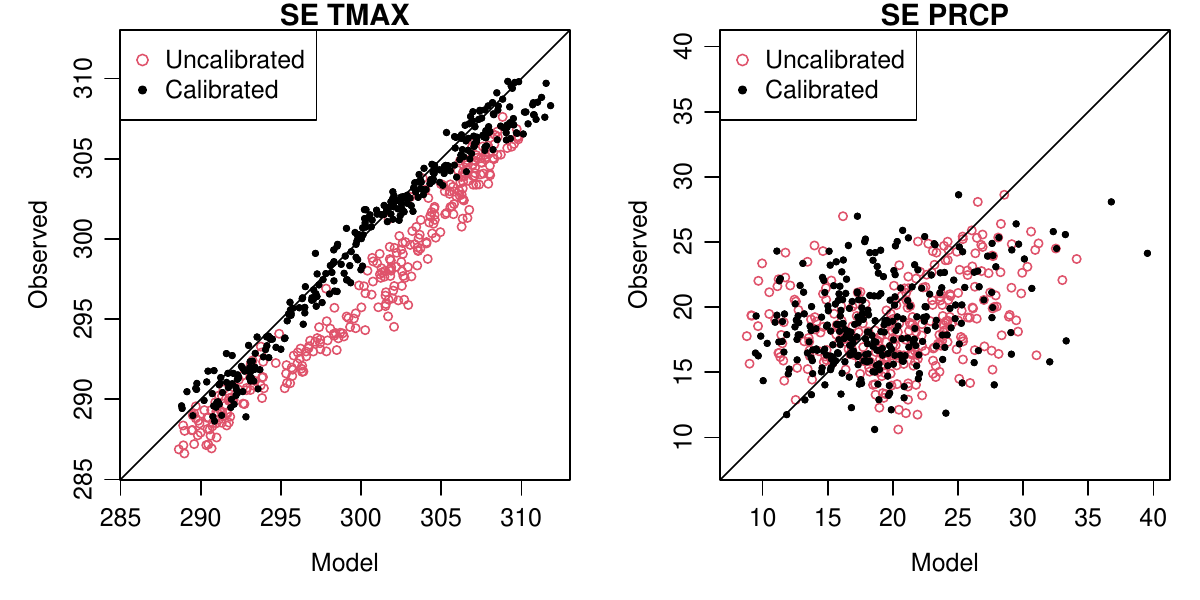}
         \caption{Monthly 0.95 quantiles for validation period of 2001--2014.}
         \label{fig:summaries_validation}
     \end{subfigure}
        \caption{Monthly 0.95 quantiles of TMAX and PRCP data in the Southwest (SW) and Southeast (SE). Each point corresponds to a particular month and location.}
        \label{fig:summaries}
\end{figure}

\begin{figure}
     \centering
     \begin{subfigure}[b]{\textwidth}
         \centering
         \includegraphics[scale=0.35]{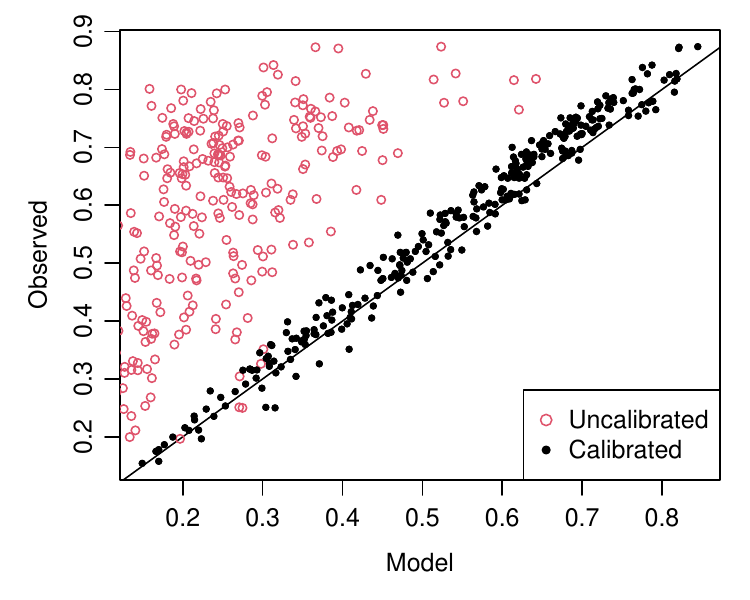}
    \includegraphics[scale=0.35]{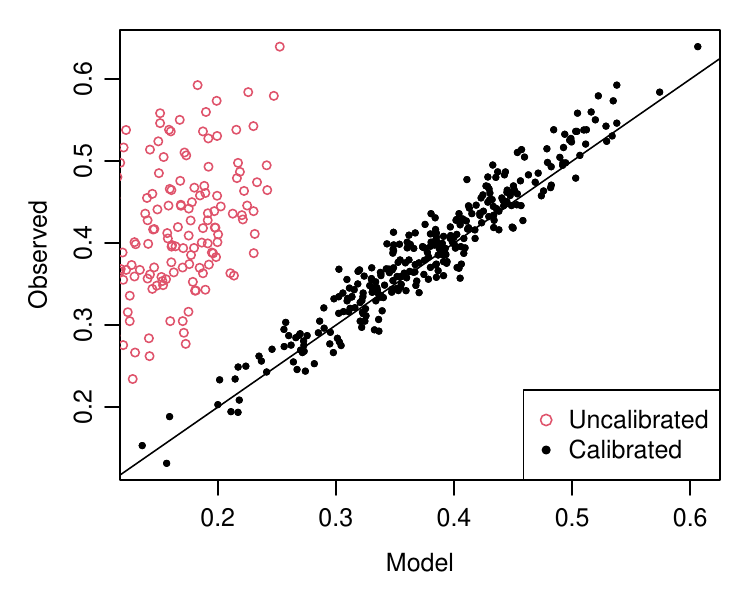}
         \caption{Proportion of zeros in training period (1951--2000) for SW (left) and SE (right).}
         \label{fig:propzero_train}
     \end{subfigure}
     \hfill
     \begin{subfigure}[b]{\textwidth}
         \centering
         \includegraphics[scale=0.35]{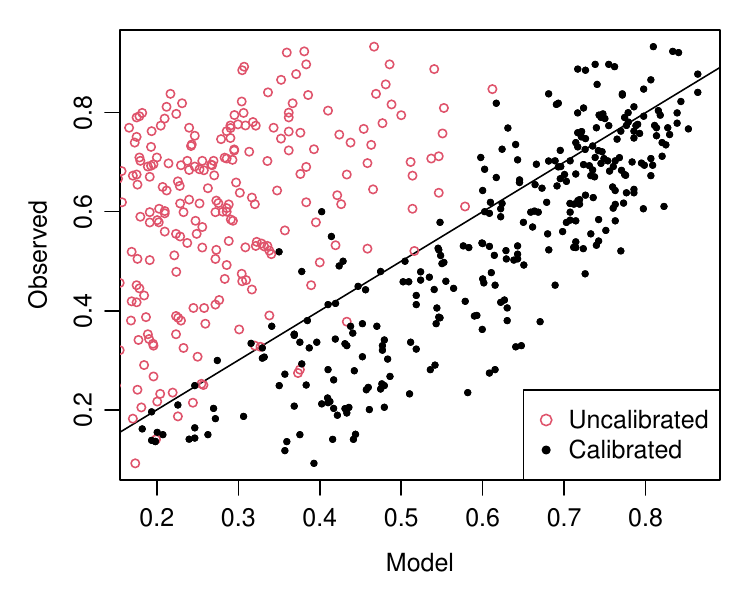}
    \includegraphics[scale=0.35]{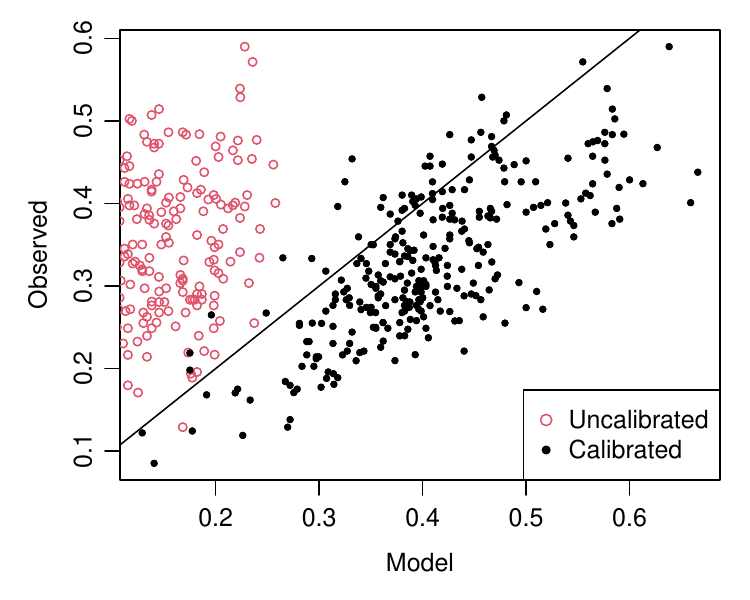}
         \caption{Proportion of zeros in validation period (2001--2014) for SW (left) and SE (right).}
         \label{fig:propzero_validation}
     \end{subfigure}
        \caption{Monthly Proportion of zeros in PRCP data for the Southwest (SW) and Southeast (SE). Each point corresponds to a particular month and location.}
        \label{fig:propzero}
\end{figure}

Figures \ref{fig:summaries}--\ref{fig:propzero} plot the 0.95 quantiles of TMAX and PRCP, and the proportion of zeros for PRCP, respectively. Each panel displays the effect of density correction by comparing it against the observational data for every location and month. For the upper 0.95 quantiles, the uncalibrated model values for TMAX have noticeable bias, something that can be visually verified in Figure \ref{fig:density}. The calibrated quantiles virtually eliminates this bias while also having lower uncertainty than the uncalibrated data, for both training and validation periods. The density correction also improves upper quantile estimation for PRCP; however, there is noticeably more uncertainty in the PRCP estimates than TMAX. The calibration is adequate for the training period, but SPQR has difficulty calibrating the upper tail of the out-of-sample PRCP for the validation period. This is consistent with the numerical study in Section \ref{s:sim}. We also note that the calibration is overall better for the SW than the SE. For estimating the proportion of zeros, we treat model precipitation below $0.001$ mm as zero precipitation, since it is not measurable by physical instruments like rain gauges. This was also done for the SPQR estimates since the continuous fitted distribution for PRCP will never produce exact zeros. The uncalibrated model data (red) in Figure \ref{fig:propzero} severely underestimates the proportion of zeros due to the drizzle effect. The calibrated density of PRCP is able to accurately represent the proportion of zeros that are observed in the real data. Calibration for the validation period (Fig \ref{fig:propzero_validation}) has some bias indicative of an over-correction and more variability compared to the training period (Fig \ref{fig:propzero_train}), but the overall results suggest that our log-transform for modeling PRCP is adequate for this application. 
\begin{figure}
     \centering
     \begin{subfigure}[b]{\textwidth}
         \centering
        \includegraphics[scale=0.35]{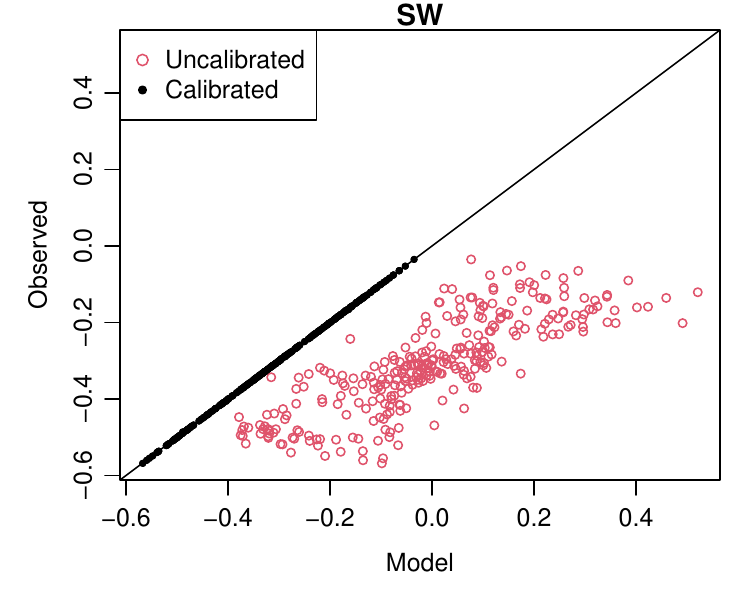}
    \includegraphics[scale=0.35]{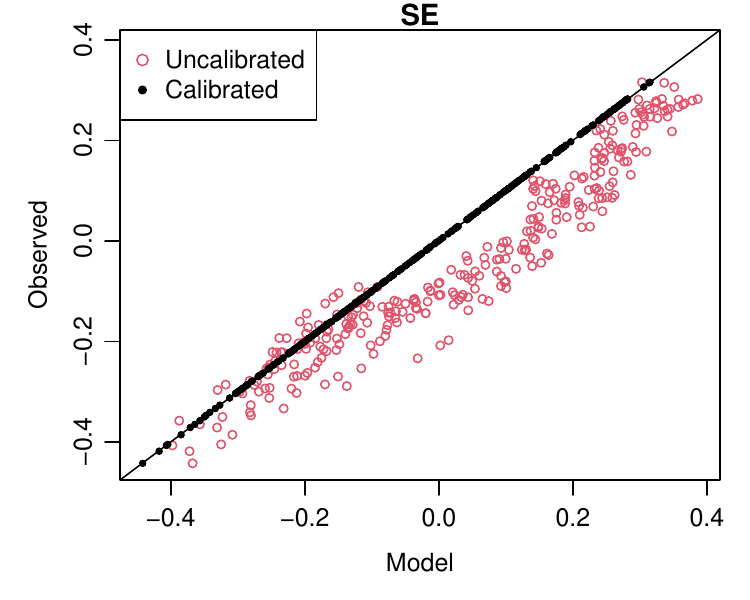}
         \caption{Cross correlations for training period of 1951--2000.}
         \label{fig:crosscorr_space_train}
     \end{subfigure}
     \hfill
     \begin{subfigure}[b]{\textwidth}
         \centering
         \includegraphics[scale=0.35]{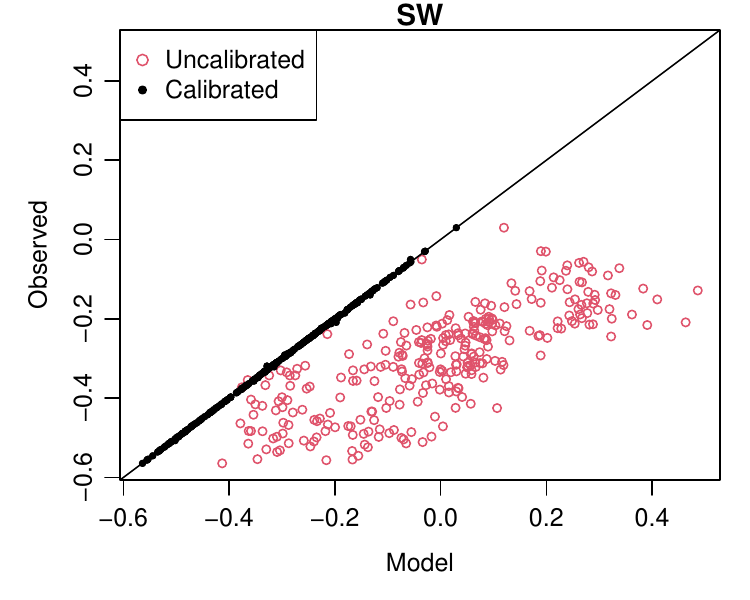}
    \includegraphics[scale=0.35]{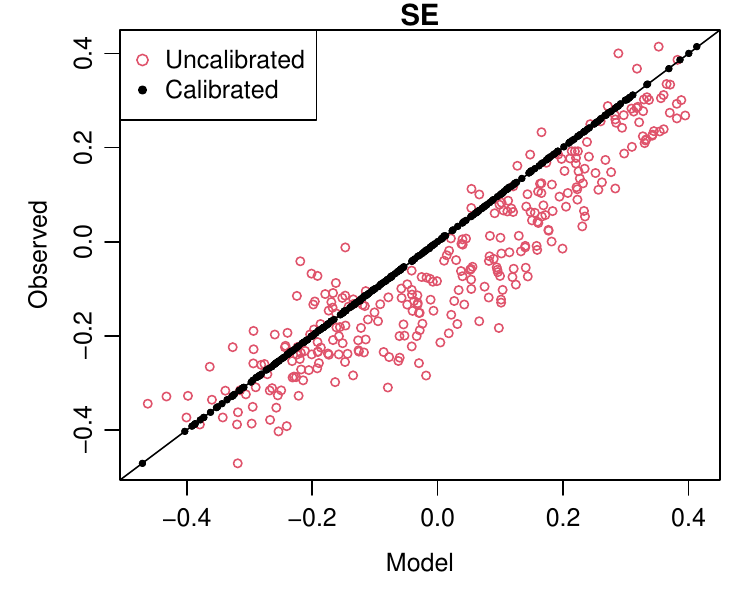}
         \caption{Cross correlations for validation period of 2001--2014.}
         \label{fig:crosscorr_space_validation}
     \end{subfigure}
        \caption{Monthly cross-correlations between TMAX and PRCP in the Southwest (SW) and Southeast (SE). Each point corresponds to a particular month and location.}
        \label{fig:crosscorr_space}
\end{figure}

\begin{figure}
     \centering
     \begin{subfigure}[b]{\textwidth}
         \centering
        \includegraphics[scale=0.35]{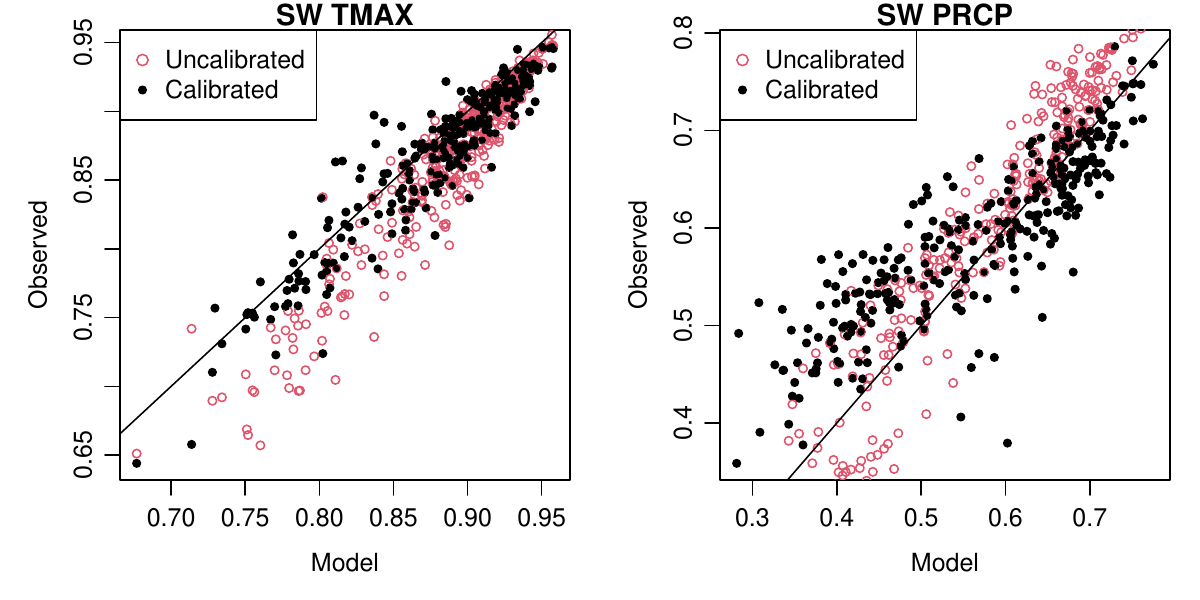}
         \includegraphics[scale=0.35]{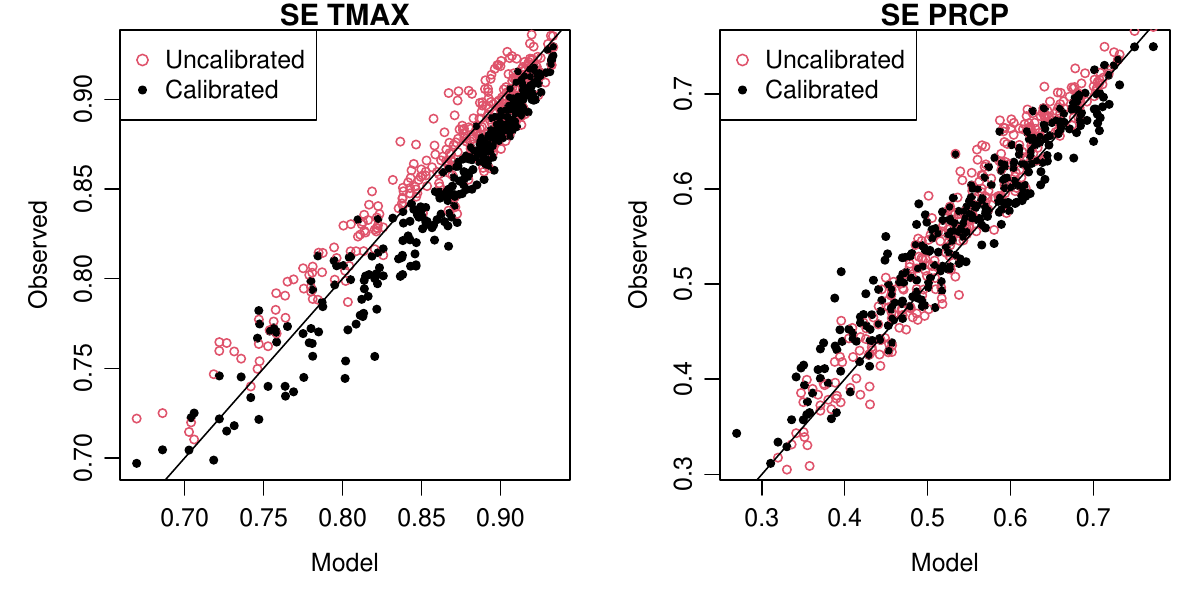}
         \caption{Spatial correlations for training period of 1951--2000.}
         \label{fig:spatcorr_space_train}
     \end{subfigure}
     \vfill
     \begin{subfigure}[b]{\textwidth}
         \centering
         \includegraphics[scale=0.35]{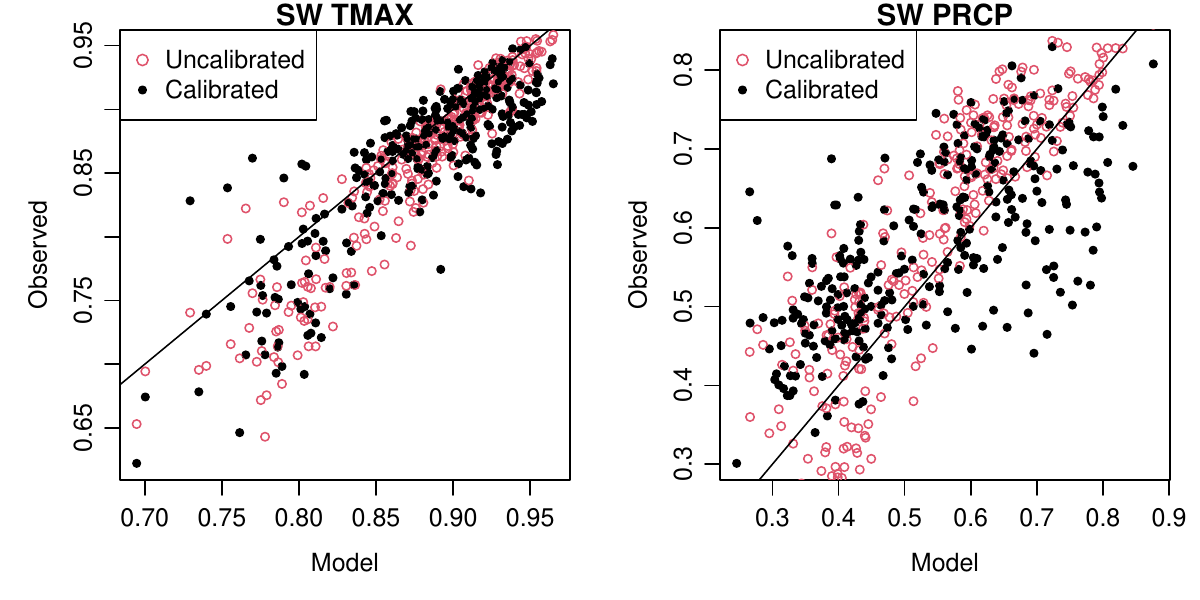}
         \includegraphics[scale=0.35]{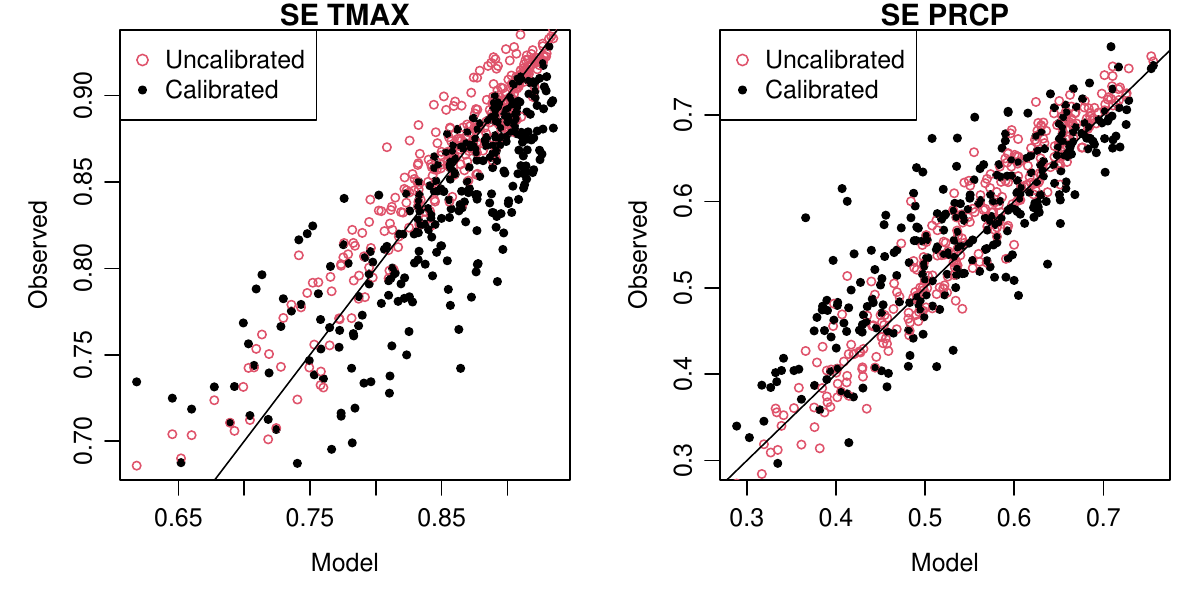}
         \caption{Spatial correlations for validation period of 2001--2014.}
         \label{fig:spatcorr_space_validation}
     \end{subfigure}
        \caption{Annual pairwise spatial correlations of TMAX and PRCP data in the Southwest (SW) and Southeast (SE). Each point corresponds to a pair of locations.}
        \label{fig:spatcorr_space}
\end{figure}

Figure \ref{fig:crosscorr_space} plots the effect of density correction on the cross correlation between variables. The uncalibrated data has positive bias in the cross correlations for both regions, with the bias in the SW being more pronounced. Density correction was able to virtually eliminate the bias and provided highly accurate and precise estimates of the cross correlation during both the training and validation periods. This is in part because the conditional density of PRCP depends on both current and past TMAX values, and therefore SPQR models the dependence between the two variables.

Figure \ref{fig:spatcorr_space} shows pairwise spatial correlations for the calibrated and uncalibrated data. Each panel contains values for 300 pairs of locations. For the training period (Fig \ref{fig:spatcorr_space_train}), values are largely concentrated around the diagonal, suggesting the the calibrated data has similar spatial correlation compared to the observed data. The uncertainty in the estimates is higher for the SW than the SE, and pairs of locations with lower spatial correlations have more uncertainty in their estimates. For the validation period (Fig \ref{fig:spatcorr_space_validation}), the uncertainties in the estimates are more pronounced. Even in such situations, the maximum error in the estimates is only $\approx 0.10$. The differences in the amount of calibration between the training and validation periods can be attributed to the dependence structure changing between the two periods and to modeling PRCP in the log-scale; additional details are provided in Appendix \ref{s:App_stationarity}. 

\begin{figure}
     \centering
     \begin{subfigure}[b]{\textwidth}
         \centering
        \includegraphics[scale=0.3]{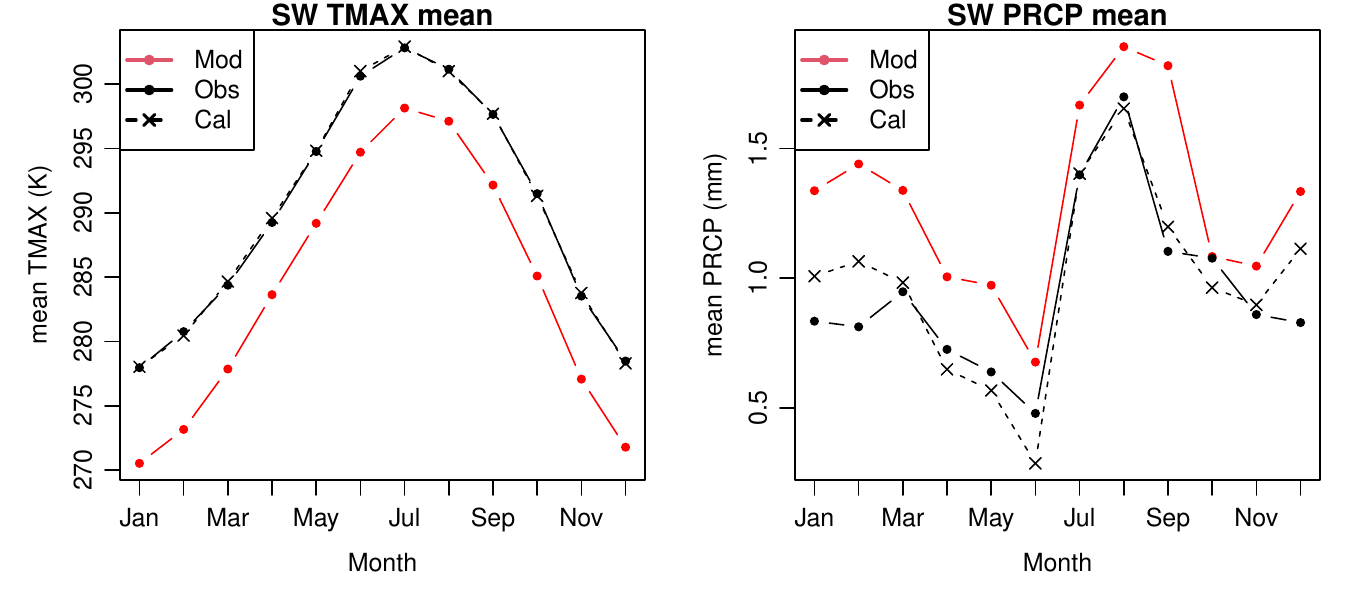}
         \includegraphics[scale=0.3]{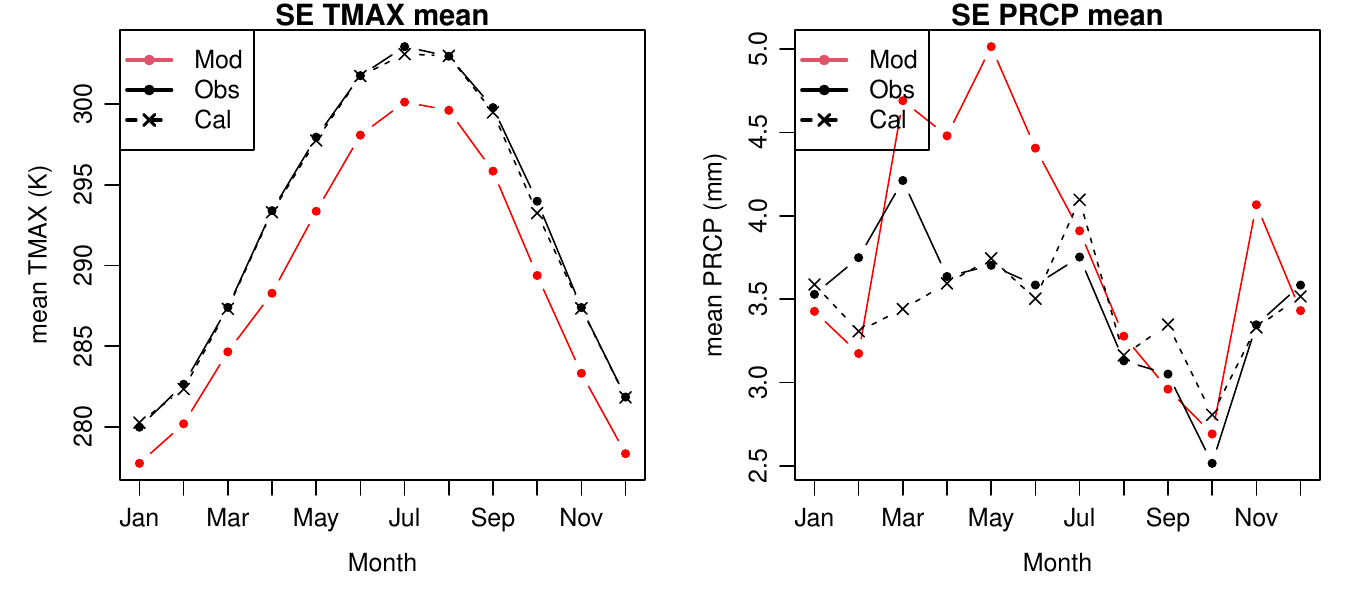}
         \caption{Monthly means for training period of 1951--2000.}
         \label{fig:means_monthly_train}
     \end{subfigure}
     \vfill
     \begin{subfigure}[b]{\textwidth}
         \centering
         \includegraphics[scale=0.3]{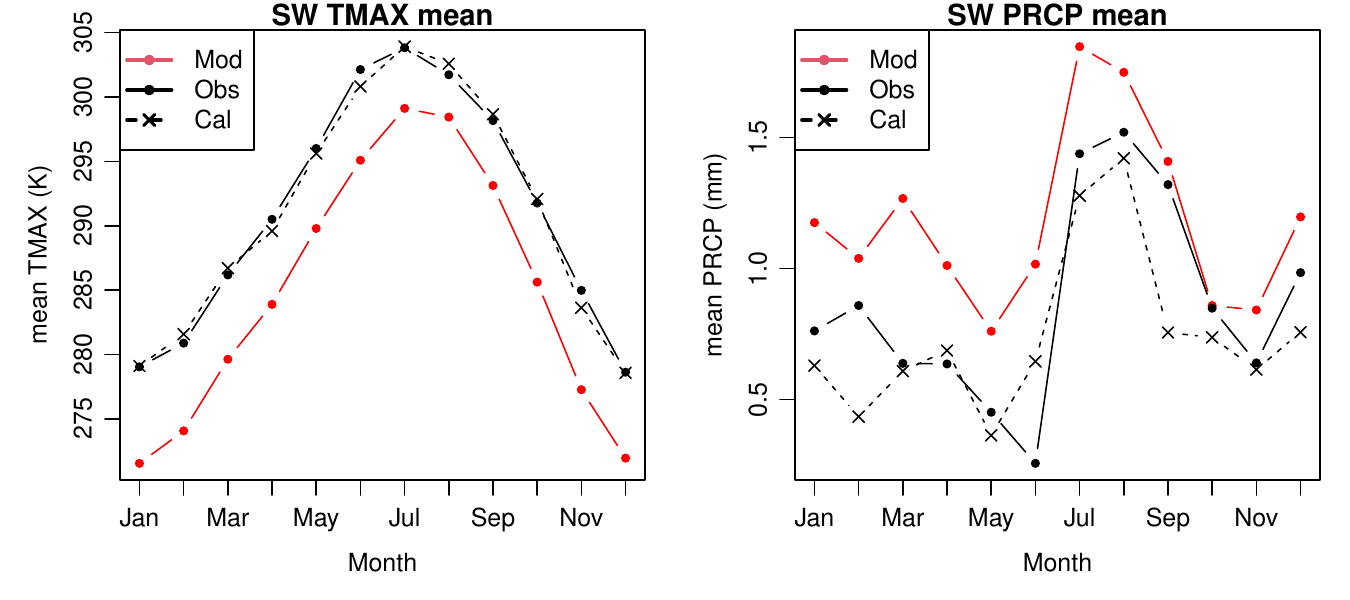}
         \includegraphics[scale=0.3]{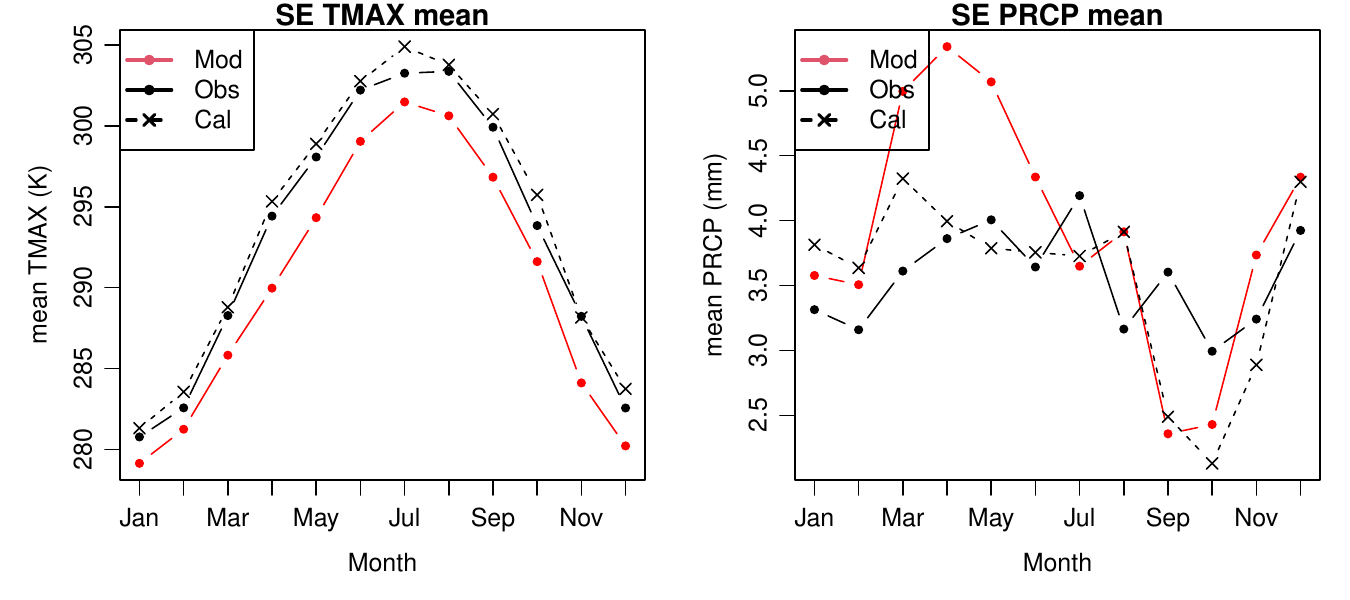}
         \caption{Monthly means for validation period of 2001--2014.}
         \label{fig:means_monthly_validation}
     \end{subfigure}
        \caption{Monthly means of TMAX and PRCP data in the Southwest (SW) and Southeast (SE), pooled across all 25 locations.}
        \label{fig:means_monthly}
\end{figure}

Finally, we also verified whether SPECD can capture monthly and seasonal characteristics of PRCP and TMAX. Figure \ref{fig:means_monthly} plots the monthly means of the two variables for each region. For both the training (Fig \ref{fig:means_monthly_train}) and validation (Fig \ref{fig:means_monthly_validation}), TMAX is calibrated very accurately, lining up with the observed data across all panels. The calibration process maps the distribution of the model data onto the distribution of the observational data; since the shape of the distribution of TMAX does not change between the two periods, the calibration works very well for the variable. The distribution of monthly PRCP, however, shows changes between the training and the validation periods. The changes occur in both the model and observational process, and are more pronounced in the SE compared to the SW. In the SW, the training period PRCP is calibrated adequately, with some deviations in the winter months of Dec-Feb. For the validation period, the calibration struggles in months where the relative behavior of model and observational data is very different from the training period. For example, Feb and Sep are two months where the model and observational data differ the most during the training period, but which are very close to each other in the validation period. A similar pattern is seen for SE PRCP. The calibration for the training period is adequate for most months, but the calibrated PRCP in the validation period is an interpolation of the training and validation period means of the observational data. As before, months where the relative behavior of the model and observed data change the most between the two periods are ones where the calibration struggles the most. Since neural networks often find generalizing outside the range of the training data challenging, changes in the joint distribution of observed and model data affect predictive accuracy of SPECD since it has not been trained on the distribution of PRCP in the validation period. Appendix \ref{s:App_seasonal} compares additional seasonal density fits, the upper quantiles of TMAX and PRCP, and proportion of zeros for PRCP. In particular, we note that though the calibration of monthly 0.95 quantiles of PRCP show the same behavior as the mean (likely due to the log-transform), the proportion of zeros are well-calibrated for both periods.
\subsection{Comparison of SPECD with CCA and QM}\label{s:app:compare}

\begin{table}
\small
    \begin{subtable}[h]{\textwidth}
        \centering
       \begin{tabular}{clcccccc}
 & \multicolumn{1}{c}{} & \multicolumn{3}{c}{\textbf{Southwest}} & \multicolumn{3}{c}{\textbf{Southeast}} \\
 & \multicolumn{1}{c}{} & SPECD & QM & CCA & SPECD & QM & CCA \\\toprule
\multirow{3}{*}{\rotatebox[origin=c]{90}{\textbf{TMAX}}} & Wass. distance & 0.120	(0.035)&	\textbf{0.106	(0.020)}&	0.122	(0.051)&	\textbf{0.125	(0.029)}&	1.239	(1.162)&	1.241	(1.163)\\
& 0.95 quantile & 0.346	(0.478)&	\textbf{0.233	(0.328)}&	0.288	(0.326)&	\textbf{0.391	(0.511)}&	1.122	(1.596)&	1.220	(1.648)\\
 & Spatial corr. &\textbf{ 0.027	(0.034)}&	0.037	(0.044)&	0.060	(0.047)&	\textbf{0.021	(0.026)}&	0.049	(0.071)&	0.052	(0.072)\\\midrule
\multirow{4}{*}{\rotatebox[origin=c]{90}{\textbf{PRCP}}} & Wass. distance & \textbf{0.075	(0.030)}&	0.154	(0.030)&	0.408	(0.107)&	\textbf{0.216	(0.097)}&	0.550	(0.112)&	0.685	(0.168)\\
& 0.95 quantile & 0.741	(0.990)&	\textbf{0.318	(0.394)}&	0.546	(0.674)&	\textbf{1.921	(2.275)}&	1.983	(2.518)&	2.096	(2.562)\\
 & Prop. of zeros & \textbf{0.031	(0.026)}&	0.186	(0.119)&	0.231	(0.128)&	\textbf{0.021	(0.023)}&	0.143	(0.107)&	0.111	(0.085)\\
 & Spatial corr. & \textbf{0.076	(0.094)}&	\textbf{0.076	(0.099)}&	0.144	(0.088)&	\textbf{0.043	(0.053)}&	0.079	(0.101)&	0.097	(0.112)\\\midrule
 & Cross corr. & \textbf{0.001	(0.001)}&	0.307	(0.110)&	0.307	(0.110)&	\textbf{0.000	(0.000)}&	0.082	(0.058)&	0.082	(0.058)\\\bottomrule
\end{tabular}
       \caption{In-sample predictive accuracy (standard errors) for training period (1951--2000).}
       \label{tab:rmse_train}
    \end{subtable}
    \hfill
    \begin{subtable}[h]{\textwidth}
        \centering
       \begin{tabular}{clcccccc}
 & \multicolumn{1}{c}{} & \multicolumn{3}{c}{\textbf{Southwest}} & \multicolumn{3}{c}{\textbf{Southeast}} \\
 & \multicolumn{1}{c}{} & SPECD & QM & CCA & SPECD & QM & CCA \\\toprule
\multirow{3}{*}{\rotatebox[origin=c]{90}{\textbf{TMAX}}} & Wass. distance & 0.328 (0.079) & \textbf{0.307 (0.063)} & 0.927 (0.096) & \textbf{0.898 (0.139)} & 1.674 (0.890) & 1.345 (1.106)
\\
& 0.95 quantile & 0.912 (1.061) & 0.946 (1.085) & \textbf{0.749 (0.777)} & \textbf{1.022 (1.103)} & 1.767 (1.814) & 1.276 (1.753)
\\
 & Spatial corr. & \textbf{0.037 (0.050)} & \textbf{0.037 (0.049)} & 0.042 (0.052) & \textbf{0.036 (0.048)} & 0.057 (0.080) & 0.058 (0.082)
\\\midrule
\multirow{3}{*}{\rotatebox[origin=c]{90}{\textbf{PRCP}}} & Wass. distance & 0.146 (0.060) & \textbf{0.126 (0.026)} & 0.543 (0.154) & \textbf{0.233 (0.050)} & 0.478 (0.143) & 0.857 (0.235)
\\
& 0.95 quantile & 1.302 (1.604) & \textbf{1.276 (1.650)} & 1.927 (1.530) & 4.040 (5.057) & 4.660 (5.791) & \textbf{4.020 (3.442)}
\\
 & Prop. of zeros & \textbf{0.103 (0.107)} & 0.171 (0.215) & 0.176 (0.175) & 0.086 (0.066) & 0.166 (0.194) & \textbf{0.084 (0.094)}
\\
 & Spatial corr. & 0.128 (0.161) & 0.117 (0.151) & \textbf{0.096 (0.124)} & \textbf{0.080 (0.101)} & 0.099 (0.129) & 0.091 (0.109)
\\\midrule
 & Cross corr. & \textbf{0.001 (0.001)} & 0.298 (0.121) & 0.298 (0.121) & \textbf{0.001 (0.001)} & 0.086 (0.079) & 0.086 (0.079)\\\bottomrule
\end{tabular}
        \caption{Out-of-sample predictive accuracy (standard errors) for validation period (2001--2014).}
        \label{tab:rmse_validation}
     \end{subtable}
     \caption{Predictive accuracy (with standard errors) in the density correction of TMAX and PRCP based on the SPECD model, compared with canonical correlation analysis (CCA) and quantile mapping (QM). Other than the Wasserstein distance metric, the remaining values correspond to the MAE between observed and calibrated data. Metrics are averaged across spatial locations (pairs of locations for spatial correlation). Best (smallest) values for each combination of region, variable, and metric are highlighted.}
     \label{tab:rmse}
     \normalsize
\end{table}

To evaluate the performance of the SPECD density correction approach compared to CCA and QM, we provide metrics for predictive accuracy based on the three methods in Table \ref{tab:rmse}. Marginal fits for TMAX and the transformed PRCP are summarized using the Wasserstein distance, while for the remainder of the metrics provided in this section, we evaluate the MAE. In each case, smaller values indicate more accurate predictions. Standard errors for the estimates are provided in parenthesis. For the training period (Table \ref{tab:rmse_train}), SPECD outperforms the competing methods for the SE across all metrics. In the more challenging SW region, QM is competitive against SPECD; it is often better at marginal metrics, whereas SPECD is better at spatial and cross correlations. For the validation period (Table \ref{tab:rmse_validation}), all three methods are competitive in specific areas. In the SE, CCA is better at calibrating the lower and upper tails of PRCP, but SPECD is better at the remaining metrics. In the SW, there is no clear winner among the three methods; while SPECD has the highest MAE for spatial correlation of PRCP, it has lower error than one or both of its competitors across the remaining metrics.

Overall, the SPECD calibrated distributions of TMAX and PRCP have the same characteristics as their observed data counterparts, for statistics of interest. In particular, it is able to very accurately capture the relationship between the two variables, which is crucial as they jointly affect several quantities of practical interest, including droughts, fire risk, and streamflow. Despite the differences in the spatial structure in the training and validation periods, it has competitive out-of-sample predictive accuracy, and excellent in-sample predictive accuracy.

\section{Discussion}\label{s:discussion}

In this paper, we develop a new conditional density estimator for the simultaneous density correction of spatial fields of PRCP and TMAX. A key goal of our study is to maintain the cross correlation between the two variables during density correction. SPECD leverages the Vecchia approximation to maintain spatiotemporal and inter-variable dependencies, and density estimation is carried out using the deep learning based SPQR method. 
Our approach is trained across two spatial fields in the Southeast and Southwest US using historical data from 1951--2000, and tested using data from 2001--2014 for the same region. SPECD is able to outperform two competing methods (QM and CCA) in most scenarios, and the joint distribution characteristics of the observational data is preserved in the calibrated data.

SPECD is highly scalable, since the Vecchia approximation allows for fitting models at different locations and months in parallel. Additionally, it ensures that the computational cost is linear in the number of grid cells. In our current implementation, however, we fit separate models for each month to capture finer seasonal effects. Future work will focus on consolidating the monthly models into a single annual model to reduce computational cost and improve estimation (since significantly more data will be available to train the annual model). We anticipate this can be achieved by including time-based covariates to account for trends and seasonality. Additionally, we plan to extend SPQR to support a bivariate response variable, which could then be used to directly model the joint distribution of TMAX and PRCP without needing to decompose it into a product of a conditional and a marginal distribution. Finally, we aim to explore the recently developed semi-parametric quantile regression for extremes framework \citep[SPQRx;][]{majumder2025SPQRx}, as a replacement for SPQR. SPQRx is designed for modeling conditional densities with heavy tails, and will likely alleviate some of the issues with modeling the upper tails of TMAX and PRCP, and being more robust to distribution shifts than SPECD based on SPQR. The ultimate goal of these anticipated extensions is to provide future projections of TMAX and PRCP across the entire contiguous US.

\section*{Data availability}
Preprocessed data and \texttt{R} Code for SPECD is available on GitHub at https://github.com/reetamm/SPECD. Additionally, the raw data as well as Python scripts for implementing the competing approaches (QM and CCE) are available on GitHub at https://github.com/fsq0511/NFbias.

\section*{Acknowledgments}

This work was supported by a grant from the National Science Foundation (DMS 2152887, CBET 2151651). Emily Hector was also supported by an Internationalization Seed Grant award from the Office of Global Engagement at North Carolina State University. 
\bibliographystyle{apalike}
\bibliography{refs}

%\clearpage
\begin{appendix}

\section{Extension of SPECD to spatiotemporal data}\label{s:mvst}
\begin{algorithm}[h]
\caption{Spatiotemporal density correction for $Y_1 = $ TMAX and $Y_2 = $ PRCP}
\label{a:alg3}
\begin{algorithmic}
    \Require Data $\bY,X$
    \For{$l = 1,\ldots,n$}
    \State $\bY_{(1lt)} = \{Y_{1l,t-1},Y_{1,l-1,t},\ldots,Y_{1,l-m,t}\}$
    \State $\bY_{(2lt)} = \{Y_{2l,t-1},Y_{1l,t-1},Y_{1lt},\ldots,Y_{1,l-m,t},Y_{2,l-1,t},\ldots,Y_{2,l-m,t}\}$
    \Procedure{SPQR for TMAX and PRCP}{}
    \State Estimate $F_{1l}$ and $Q_{1l}$ for $Y_{1lt}|\bY_{(1lt)},X$ \Comment{Estimation step}
    \State Estimate $F_{2l}$ and $Q_{2l}$ for $Y_{2lt}|\bY_{(2lt)},X$ \Comment{Estimation step}
    \EndProcedure
    
    \Procedure{TMAX density correction}{} 
    \State $u_{1lt} \gets F_{1l}(Y_{1lt}|\bY_{(1lt)},X=0)$ \Comment{Projection step}
    \For{$t = 2,\ldots,T$}
    \State $\bY_{(1lt)}^* = \{Y_{1l,t-1}^*,Y_{1,l-1,t}^*,\ldots,Y_{1,l-m,t}^*\}$
    \State $Y_{1lt}^* \gets Q_{1l}(u_{1lt}|\bY_{(1lt)}^*,X=1)$ \Comment{Calibration step}
    \EndFor
    \EndProcedure
    \Procedure{PRCP density correction}{}
    \State $u_{2lt} \gets F_{2l}(Y_{2lt}|\bY_{(2lt)},X=0)$ \Comment{Projection step}
    \For{t = 2,\ldots,T}
    \State $\bY_{(2lt)}^* = \{Y_{2l,t-1}^*,Y_{1l,t-1}^*,Y_{1lt}^*,\ldots,Y_{1,l-m,t}^*,Y_{2,l-1,t}^*,\ldots,Y_{2,l-m,t}^*\}$
    \State $Y_{2lt}^* \gets Q_{2l}(u_{2lt}|\bY_{(2lt)}^*,X=1)$ \Comment{Calibration step}
    \EndFor
    \EndProcedure
     \EndFor
\end{algorithmic}
\end{algorithm}

We extend the model from Section \ref{s:mvs} in the main text to incorporate temporal information. 
Let $Y_{1lt}$ and $Y_{2lt}$ denote the response for TMAX and PRCP at location $l \in \{1,\ldots,n\}$ and time $t\in\{1,...,T\}$. We order the $p=2 n T$ variables as $(Y_{111},\ldots,Y_{11T},Y_{211},\ldots,Y_{21T},\ldots,Y_{1n1},\ldots,Y_{1nT},\\Y_{2n1},\ldots,Y_{2nT})$. 
The SPECD model applies to multivariate time series data after key assumptions of Markovian and stationarity dependence structure.  The Markovian structure states that the conditional quantile function for $Y_{jlt},j=1,2,$ given all observation before time $t$ depends only on data from the previous $T' < T$ time steps; the stationarity assumption states that the conditional quantile functions are invariant to $t$. 

We assume the conditional quantile function to be first order Markov in time, i.e., $T'=1$. This assumption is reasonable given the selected are from the Southeast and Southwest US in regions where pronounced seasonality in precipitation is absent \citep{PETERSEN2012}. This implies that the Vecchia neighbor set for $Y_{1lt}$ is $\bY_{(1lt)} = \{Y_{1l,t-1},Y_{1,l-1,t},\ldots,Y_{11t}\}$, and for $Y_{2lt}$ is $\bY_{(2lt)} = \{Y_{2l,t-1},Y_{1l,t-1},Y_{1,l,t},\ldots, Y_{1,l-m,t},Y_{2,l-1,t},\ldots,Y_{2,l-m,t}\}$. These are straightforward extensions of the Vecchia neighbor sets presented in Section \ref{s:mvs}, with TMAX on day $t$ being additionally depending on TMAX on day $t-1$, and PRCP on day $t$ additionally depending on both PRCP and TMAX from day $t-1$.  
%The conditional quantile function for variable $j$ is 
%$$Y_{jt} = Q_j(U_{jt};\bY_{(jt)},X)$$
%for $U_{jt}\iid\mbox{Uniform}(0,1)$. 
%Everything else is the same as before.
The conditional QF $Q$, CDF $F$, and calibration function $C$, for $\bY_{lt}$ are defined identically to Section \ref{s:mv}; these do not depend on $t$ due to the stationarity assumption. Algorithm \ref{a:alg3} details the density correction methodology for spatiotemporal data.

\section{QM and CCA implementation details}\label{s:qmcca}

This section briefly outlines QM and CCA methodology that is employed to bias correct the GCM data. We refer readers to \citet{bhowmick2017} and \citet{SEO2019304} for further details on bias correction using QM and CCA. 

\subsection{Quantile mapping} 
Our QM implementation first sorts each time series (GCM and observation) individually in ascending order. It then applies a simple linear regression on the sorted data, with the observational data serving as the response, and the GCM data the covariate. Once the linear regression models have been fitted for each variable at each grid point, bias correction is carried out by substituting the raw (unsorted) GCM time series into the fitted regression equation. The steps are outlined in Algorithm \ref{a:alg_qm}.

\begin{algorithm}[h]
\caption{Quantile mapping for TMAX and PRCP}
\label{a:alg_qm}
\begin{algorithmic}
    \Require Observational TMAX ($\bY_1$) and PRCP ($\bY_2$) data 
    \Require GCM TMAX ($\bX_1$) and PRCP ($\bX_2$) data 
    \For{ location $l = 1,\ldots,n$}
    \Procedure{Asynchronous regression for TMAX and PRCP}{}
    \State Obtain $Y_{1l}' , Y_{2l}', X_{1l}',X_{2l}'$ by sorting $Y_{1l} , Y_{2l}, X_{1l},X_{2l}$ individually \Comment{Sorting step}
    \State Fit simple linear regression models $Y_{1l}'\sim X_{1l}'$ and $Y_{2l'}\sim X_{2l}'$ \Comment{QM step}
    \EndProcedure
    
    \Procedure{TMAX and PRCP bias correction}{} 
    \State Obtain predictions $\hat{Y}_{1l}$ and $\hat{Y}_{2l}$ from fitted models using raw GCM data\Comment{Bias correction step}
    \EndProcedure
     \EndFor
\end{algorithmic}
\end{algorithm}

\subsection{Canonical correlation analysis}
Canonical correlation is employed in bias correction to address one of the common issues that arise with asynchronous regression. Assume $\bY_l$ is a bivariate vector of observed TMAX and PRCP at location $l$. QM orders each component of this bivariate vector independent of the other; a similar operation is carried out for $\bX_l$, the bivariate vector of GCM data. This affects the paired nature of the two meteorological variables, and there is no guarantee that the cross-correlation between TMAX and PRCP will be preserved in the bias-corrected output. Ideally, the bivariate vectors should be ordered in a manner that does not break the pairs, and bias correction carried out afterwards.

Our CCA approach begins by fitting separate bivariate Normal distributions to $\bY_l$ and $\bX_l$. Each bivariate vector is then ordered based on their joint CDF, ensuring that (TMAX, PRCP) remains paired afterwards. The sorted variables are standardized using mean-variance transformations for each marginal distribution. CCA is carried out on the standardized data, $\bY_l'$ and $\bX_l'$:

\begin{align*}
    U_l &= \bX_{l}' A_l,\\
    V_l &= \bY_{l}' B_l,
\end{align*}
Where $U_l,V_l$ are rotated components of $\bY_l, \bX_l$ with maximum (canonical) correlation $R_l$, and $A_l$ and $B_l$ are the associated eigenvectors. The predicted $\hat{U}_l$ is obtained by plugging in $\bX_l'$, and predictions of $\hat{V}_l$ is obtained as follows:
$$\hat{V}_l = R_l + (\hat{U}_l + \sqrt{1-R_l^2}).$$ 
Next, we take the component-wise reciprocal of $B_l$ via an operation denoted as inv($B_l$), and use it to obtain a bias-corrected bivariate vector:
$$\hat{\bY}_l = \hat{V}_l\cdot \mbox{inv}(B_l).$$

Finally, the components of $\hat{\bY}_l$ are rescaled to their original units, and the sorting is reversed to obtain a time series of bias-corrected TMAX and PRCP. 

\begin{algorithm}[h]
\caption{Canonical correlation analysis for TMAX and PRCP}
\label{a:alg_cca}
\begin{algorithmic}
    \Require Observational TMAX ($\bY_1$) and PRCP ($\bY_2$) data, $\bY = (\bY_1, \bY_2)^T$
    \Require GCM TMAX ($\bX_1$) and PRCP ($\bX_2$) data, $\bX = (\bX_1,\bX_2)^T$
    \For{ location $l = 1,\ldots,n$}
    
    \Procedure{Asynchronous regression for TMAX and PRCP}{}
    \State Fit bivariate Normal distribution to $\bX_{l}$ and sort by joint CDF 
    \Comment{Sorting step}
    \State Fit bivariate Normal distribution to $\bY_{l}$ and sort by joint CDF \Comment{Sorting step}
    \State Standardize $\bX_l$ and $\bY_l$ marginally to obtain $\bX_l'$ and $\bY_l'$
    \State Apply CCA to obtain rotated components $U_l$ and $V_l$, eigenvectors $A_l$ and $B_l$, and $R_l$, the canonical correlation between $U_l$ and $V_l$ \Comment{CCA step}
    \EndProcedure
    
        \Procedure{TMAX and PRCP bias correction}{} 
    \State Obtain sorted predictions for the rotated components:
    \State $\hat{U}_{l} \gets \bX_{l}'A_{l}$
    \State $\hat{V}_{l} \gets R_{l} + \hat{U}_{l}\sqrt{1+R_l^2}$
    \State Obtain sorted and standardized predictions $\hat{\bY}_{l}' \gets \hat{V}_{l}\cdot \mbox{inv}(B_{l})$ \Comment{Bias correction step}
    \State Scale and unsort to obtain bias-corrected daily records $\hat{\bY}_{l}$
    \EndProcedure
     \EndFor
\end{algorithmic}
\end{algorithm}

\section{Additional numerical study results}\label{s:AppB_sim}
This section contains additional results for the numerical study presented in Section \ref{s:sim} of the main text. Table \ref{tab:rmse_sim} describes the numerical results corresponding to Figure \ref{fig:sim_results}. Each value in the tables corresponds to a metric averaged over 100 replicates of the simulation study. Values in the parentheses represent the standard errors of the estimates.

\begin{table}[h]
\small
    \begin{subtable}[h]{\textwidth}
        \centering
       \begin{tabular}{cccccc}
 &  & QM & CCE & SPECD1 & SPECD2 \\\toprule
\multirow{3}{*}{\rotatebox[origin=c]{90}{\textbf{TMAX}}} & Wass. distance & 1.7888 (0.1267) & 2.0463 (0.5496) & \textbf{0.0993 (0.0112)} & 0.2697 (0.0216) \\
 & 0.95 quantile & 1.811 (0.1339) & 2.3730 (0.7139) & \textbf{0.1320 (0.0328)} & 0.5840 (0.0589) \\
 & Spatial corr. & 0.4831 (0.0098) & 0.7987 (0.0072) & 0.4723 (0.0112) & \textbf{0.1411 (0.0208)} \\\midrule
\multirow{4}{*}{\rotatebox[origin=c]{90}{\textbf{PRCP}}} & Wass. distance & 0.1037 (0.0034) & 0.1123 (0.0014) & \textbf{0.0265 (0.0018)} & 0.0680 (0.0080) \\
 & 0.95 quantile & 0.1819 (0.0167) & 0.1028 (0.0167) & \textbf{0.0725 (0.0132)} & 0.2462 (0.0693) \\
 & Spatial corr. & 0.4770 (0.0123) & 0.8357 (0.0107) & 0.3787 (0.0230) & \textbf{0.2607 (0.0199)} \\
 & Prop. of zeros & 0.2050 (0.0040) & 0.2069 (0.0020) & \textbf{0.0261 (0.0036)} & 0.0363 (0.0077) \\\midrule
 & Cross corr. & 0.3364 (0.0195) & 0.3031 (0.0052) & 0.0005 (2.89e-05) & \textbf{0.0004 (2.39e-04)}\\\bottomrule
\end{tabular}
       \caption{In-sample predictive accuracy (standard errors) for training period (1951--2000).}
       \label{tab:rmse_sim_train}
    \end{subtable}
    \hfill
    \begin{subtable}[h]{\textwidth}
        \centering
       \begin{tabular}{cccccc}
 &  & QM & CCE & SPECD1 & SPECD2 \\\toprule
\multirow{3}{*}{\rotatebox[origin=c]{90}{\textbf{TMAX}}} & Wass. distance & 1.7917 (0.1414) & 2.0812 (0.5655) & \textbf{0.2681 (0.0436)} & 0.3552 (0.0378) \\
 & 0.95 quantile & 1.8215 (0.1632) & 2.3480 (0.7162) & \textbf{0.3398 (0.0749)} & 0.6236 (0.1232) \\
 & Spatial corr. & 0.4839 (0.0193) & 0.7955 (0.0164) & 0.4728 (0.0212) & \textbf{0.1517 (0.0291)} \\\midrule
\multirow{4}{*}{\rotatebox[origin=c]{90}{\textbf{PRCP}}} & Wass. distance & 0.1067 (0.0089) & 0.0785 (0.0033) & \textbf{0.0471 (0.0089)} & 0.0801 (0.0175) \\
 & 0.95 quantile & 0.1894 (0.0395) & 0.1381 (0.0332) & \textbf{0.1375 (0.0321)} & 0.2949 (0.1060) \\
 & Spatial corr. & 0.4750 (0.0273) & 0.8326 (0.0246) & 0.3970 (0.0324) & \textbf{0.2514 (0.0459)} \\
 & Prop. of zeros & 0.2041 (0.0151) & 0.1683 (0.0070) & \textbf{0.0349 (0.0108)} & 0.0393 (0.0101) \\\midrule
 & Cross corr. & 0.3376 (0.0326) & 0.4032 (0.0113) & \textbf{0.0007 (0.0003)} & 0.0012 (0.0007)\\\bottomrule
\end{tabular}
        \caption{Out-of-sample predictive accuracy (standard errors) for validation period (2001--2014).}
        \label{tab:rmse_sim_test}
     \end{subtable}
     \caption{Predictive accuracy estimates (standard errors) in the density correction of TMAX and PRCP based on QM, CCA, and SPECD, based on a simulation study with 100 replicates. Other than the Wasserstein distance metric, the remaining values correspond to the MAE between observed and calibrated data. Metrics are averaged across spatial locations (pairs of locations for spatial correlation). Best (smallest) values for each combination of region, variable, and metric are highlighted.}
     \label{tab:rmse_sim}
     \normalsize
\end{table}

\clearpage

\section{Non-stationarity in the spatial dependence}\label{s:App_stationarity}
\begin{figure}
    \centering
    \includegraphics[width=\linewidth]{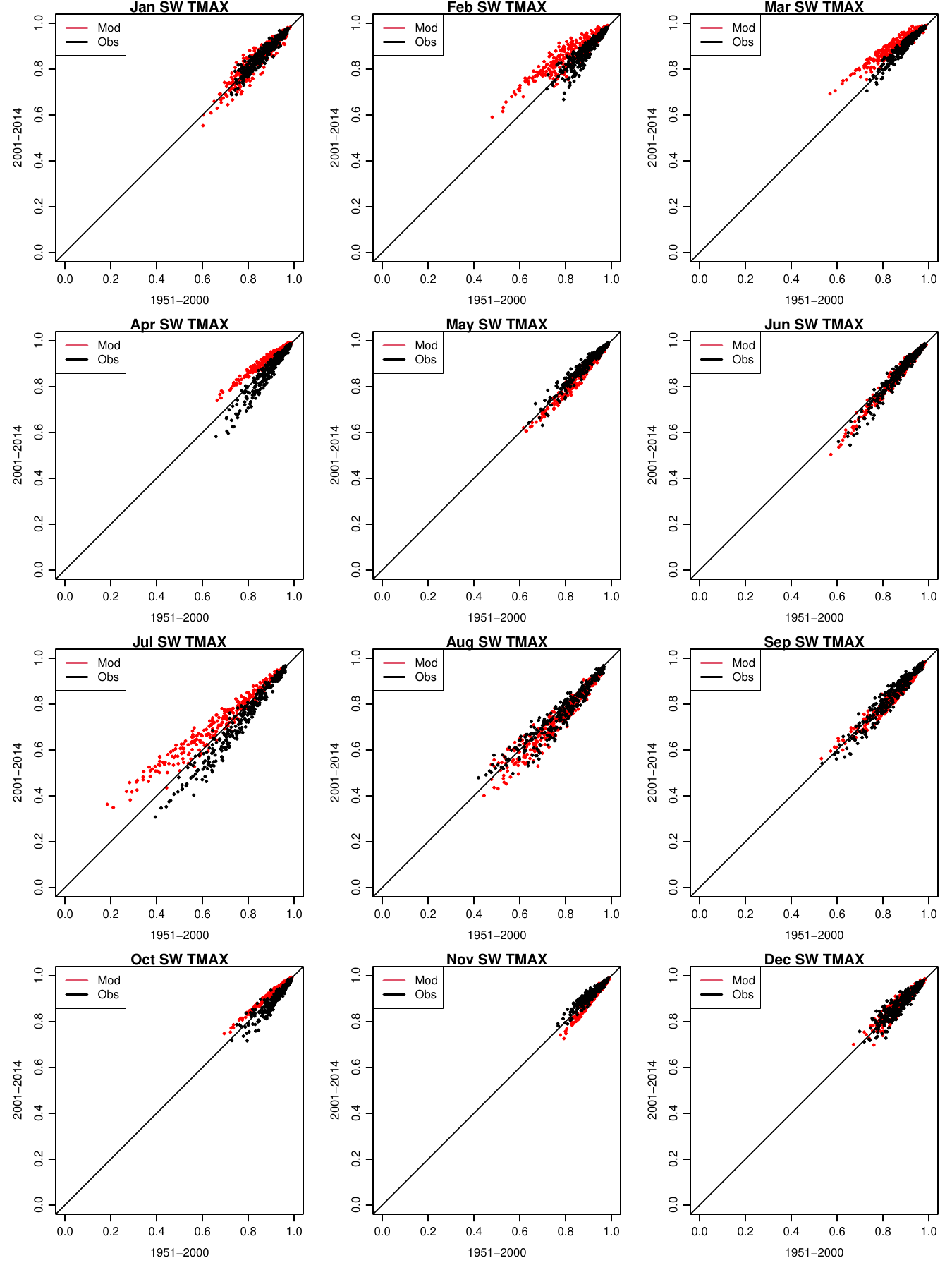}
    \caption{Comparison of pairwise spatial correlations for TMAX in the SW region between the training (1951--2000) and validation (2001--2014) periods.}
    \label{fig:dist_shift_SW_TMAX}
\end{figure}

\begin{figure}
    \centering
    \includegraphics[width=\linewidth]{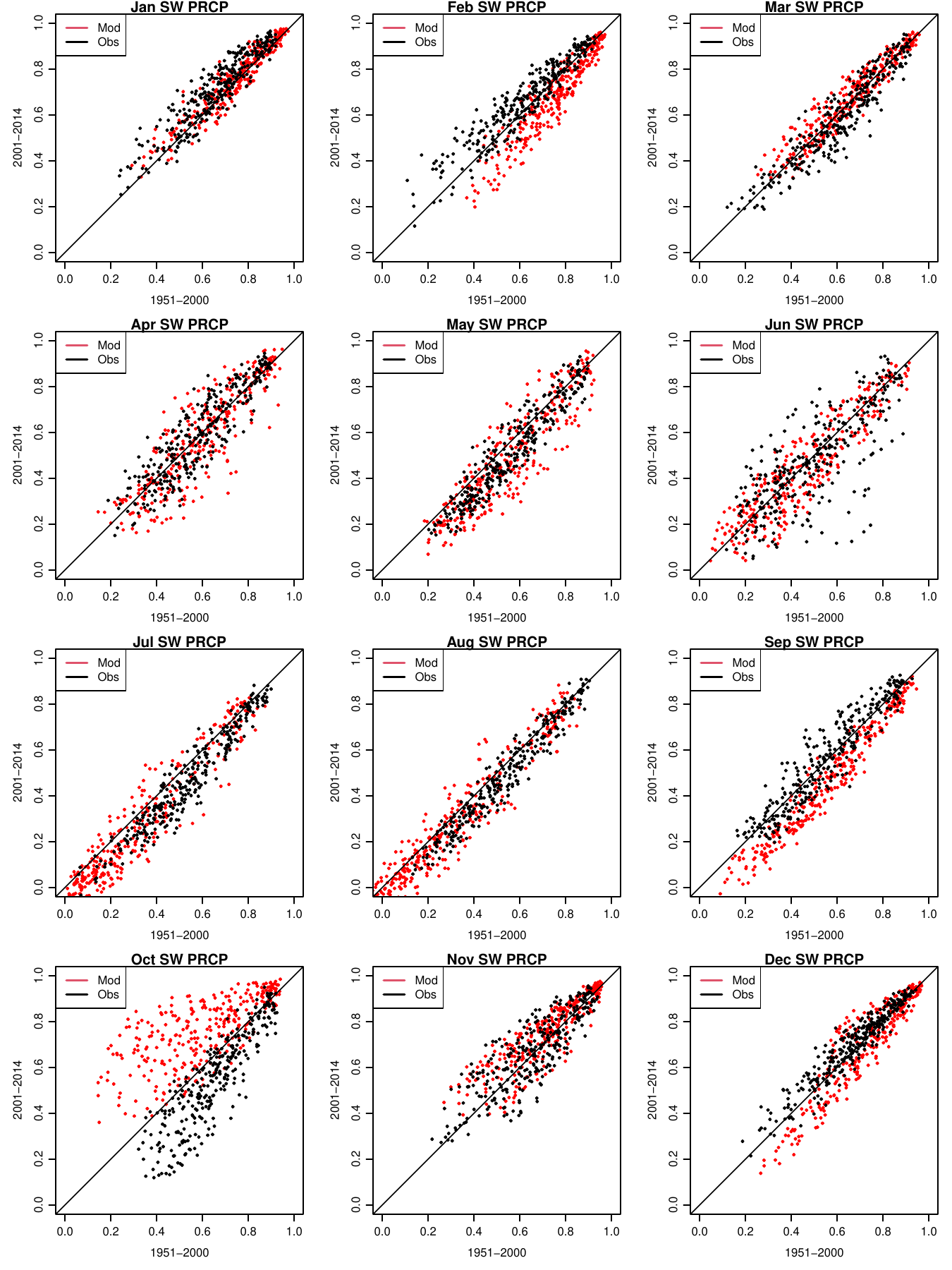}
    \caption{Comparison of pairwise spatial correlations for PRCP in the SW region between the training (1951--2000) and validation (2001--2014) periods.}
    \label{fig:dist_shift_SW_PRCP}
\end{figure}

\begin{figure}
    \centering
    \includegraphics[width=\linewidth]{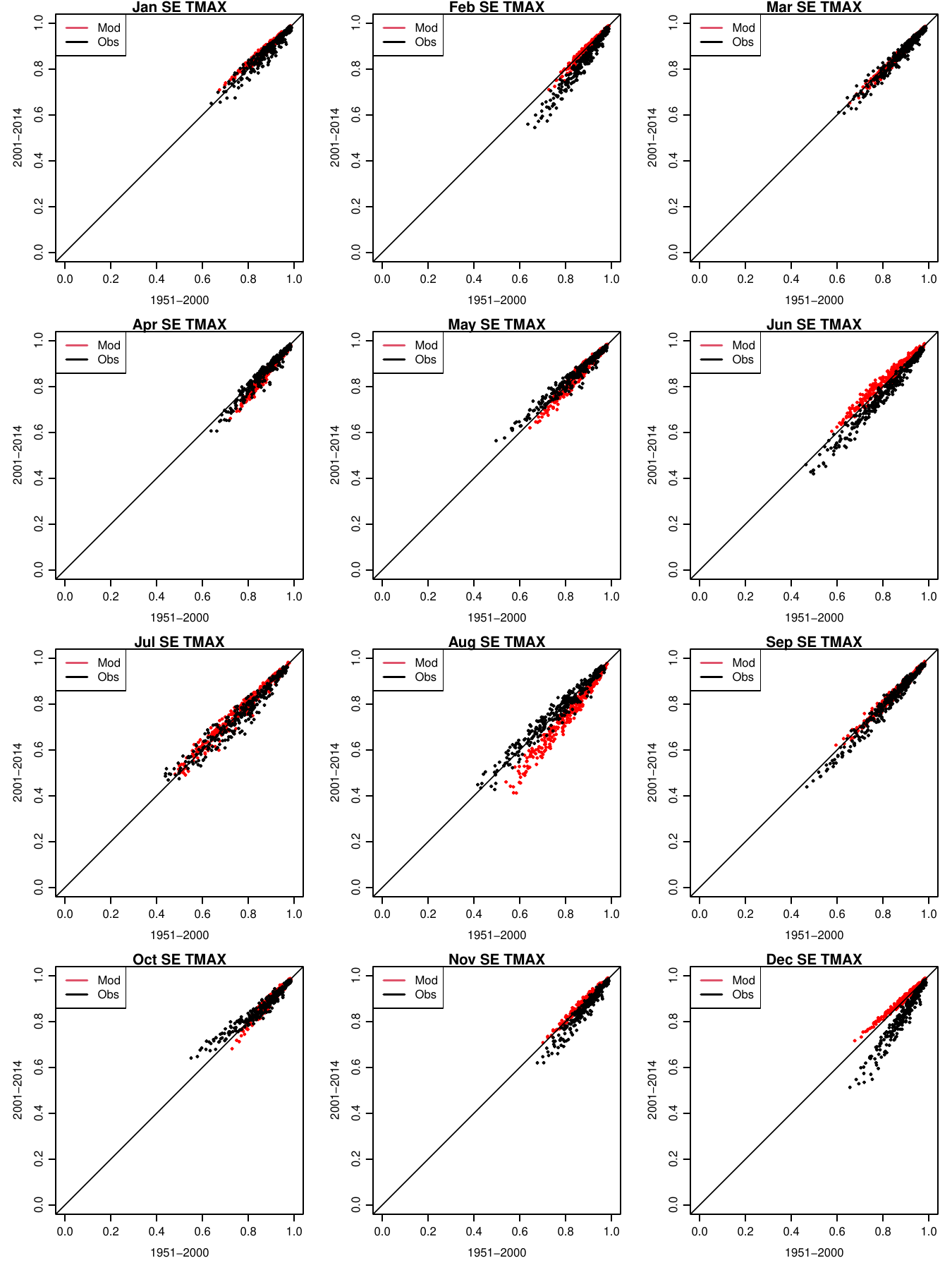}
    \caption{Comparison of pairwise spatial correlations for TMAX in the SE region between the training (1951--2000) and validation (2001--2014) periods.}
    \label{fig:dist_shift_SE_TMAX}
\end{figure}

\begin{figure}
    \centering
    \includegraphics[width=\linewidth]{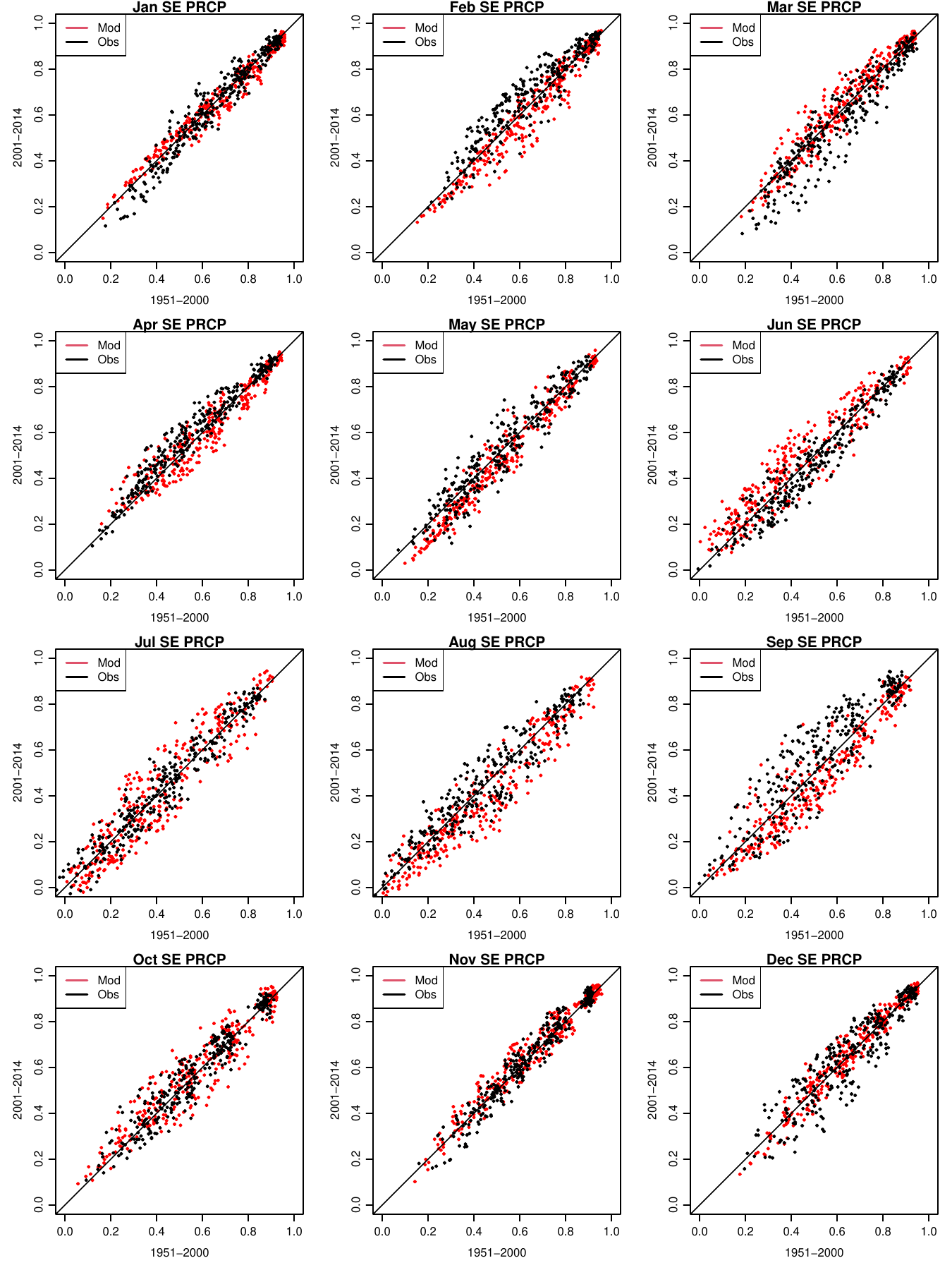}
    \caption{Comparison of pairwise spatial correlations for PRCP in the SE region between the training (1951--2000) and validation (2001--2014) periods.}
    \label{fig:dist_shift_SE_PRCP}
\end{figure}

In this section, we aim to study a shift in the spatial correlations between the training and validation periods of our study. We achieve this by comparing the pairwise spatial correlations for each combination of month, variable, region, and data source (model vs. observed). Ideally, for any of these combinations, we would like the spatial distribution to remain unchanged over time if we want to accurately assess the out-of-sample performance of the density correction methods. Figures \ref{fig:dist_shift_SW_TMAX}--\ref{fig:dist_shift_SE_PRCP} plot the pairwise spatial correlations with the training and validation periods on the two axes. Model data is represented in red, and observational data is represented in black. Deviation of points from the diagonal could indicate one of the following (non-exhaustive) scenarios:
\begin{enumerate}
    \item[1.] \textbf{Low variability, bias from one source:} Both data sources show low variability in the their estimates for the two periods, but one of the data sources shows bias; e.g., the spatial correlations in the training period is higher/lower than the spatial correlations in the validation period.
    \item[2.] \textbf{Low variability, bias in both sources:} Both data sources show low variability and both of them show bias. The bias may be in the same or opposite directions; e.g., the correlations in the model data increase while those in the observational data decrease.
    \item[3.] \textbf{High variability, low bias:} Both data sources show high variability, i.e., the spatial correlation estimates differ noticeably in the two time periods. However, bias is low, i.e., the differences in the spatial correlations from the two time periods are centered at zero.
    \item[4.] \textbf{High variability, bias from at least one source:} Both data sources show high variability in their estimates. Additionally, at least one of the variables has bias. If both variables have bias, they may be in the same direction or in opposite directions.
\end{enumerate}
Figures \ref{fig:dist_shift_SW_TMAX} and \ref{fig:dist_shift_SE_TMAX} plot the spatial correlations in TMAX for the SW and SE regions. These plots have low variability, and largely low bias. Some of the months correspond to Scenario 1 (e.g., August and December plots for the SE) or Scenario 2 (e.g., April and July in the SW).

Figures \ref{fig:dist_shift_SW_PRCP} and \ref{fig:dist_shift_SE_PRCP} plot the spatial correlations in PRCP for the SW and SE regions. These plots have higher variability, likely due to the presence of zeros in the data. Additionally, the SW has higher variability than the SE. Plots for the SE mostly correspond to Scenario 3, but some plots for the SW do demonstrate behavior consistent with Scenario 4 (e.g., February and October). 

Overall, we also notice more deviations from the diagonal that correspond to relatively low correlation values, and stronger agreement when the correlations are the highest. These results affect the calibration based on SPECD across all its metrics, not just the spatial correlations. This is because spatial information is encoded within the model by the Vecchia approximation, whereby values at nearby grid cells are used as covariates. In particular, we would expect to see better calibration for TMAX than for PRCP, and for the SE compared to the SW.

\clearpage
\section{Additional seasonal results}\label{s:App_seasonal}
\begin{figure}
     \centering
     \begin{subfigure}[b]{\textwidth}
         \centering
        \includegraphics[scale=0.45]{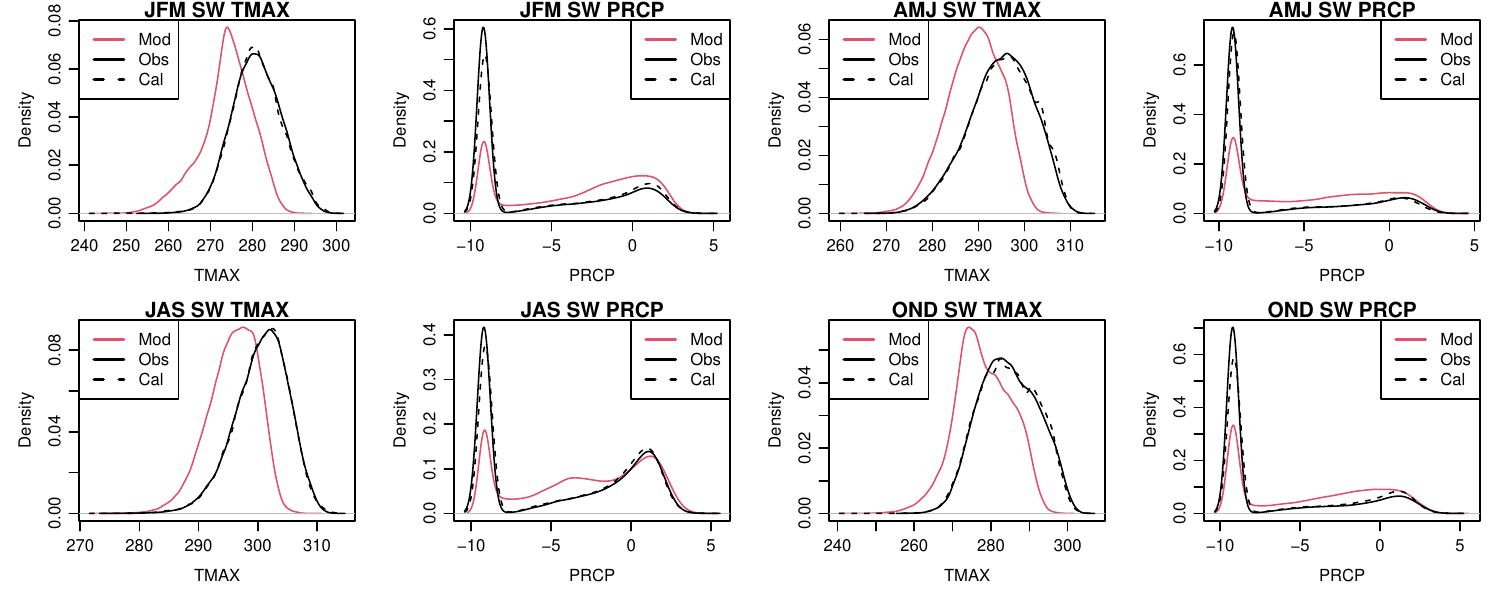}
         \includegraphics[scale=0.45]{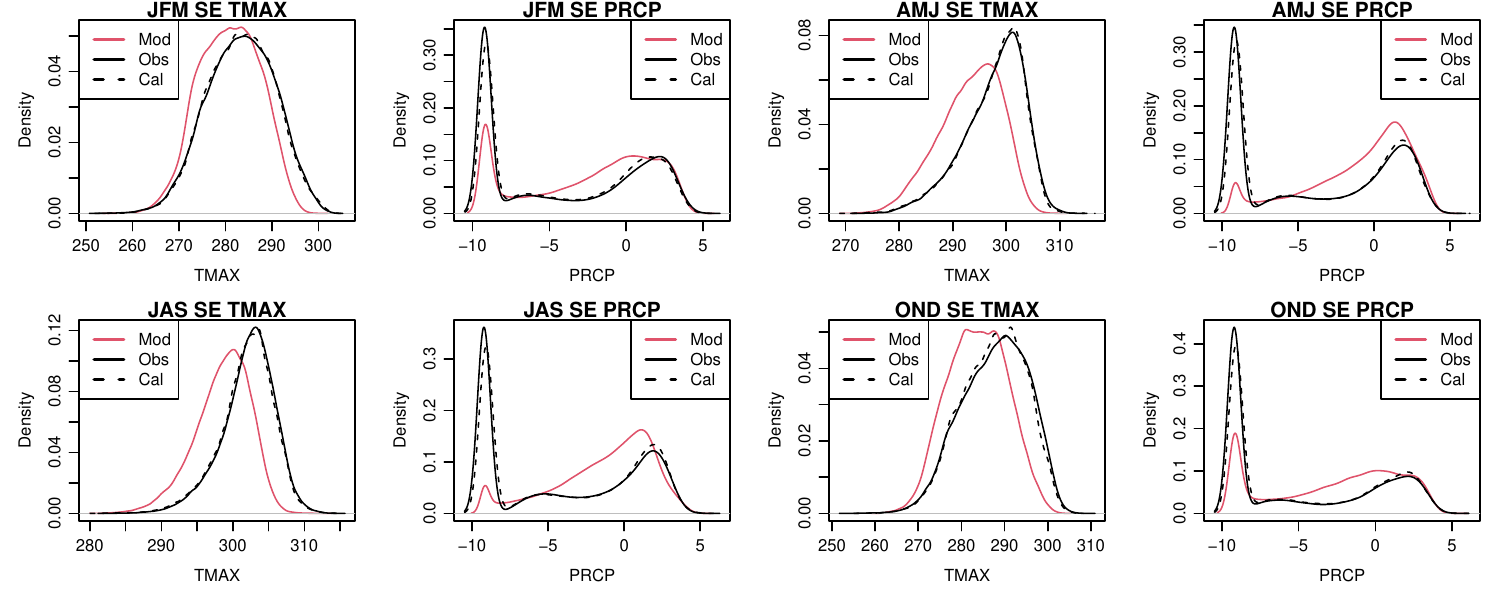}
         \caption{Seasonal density plots for training period of 1951--2000.}
         \label{fig:density_seasonal_train}
     \end{subfigure}
     \vfill
     \begin{subfigure}[b]{\textwidth}
         \centering
         \includegraphics[scale=0.45]{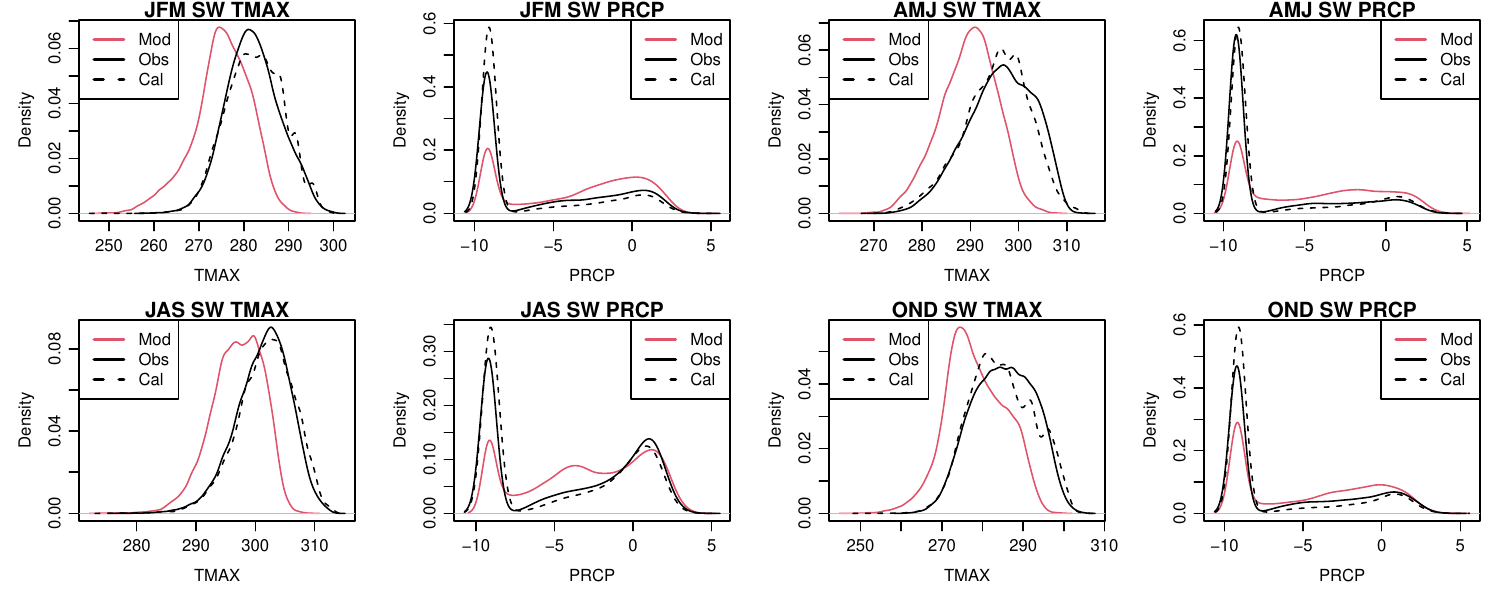}
         \includegraphics[scale=0.45]{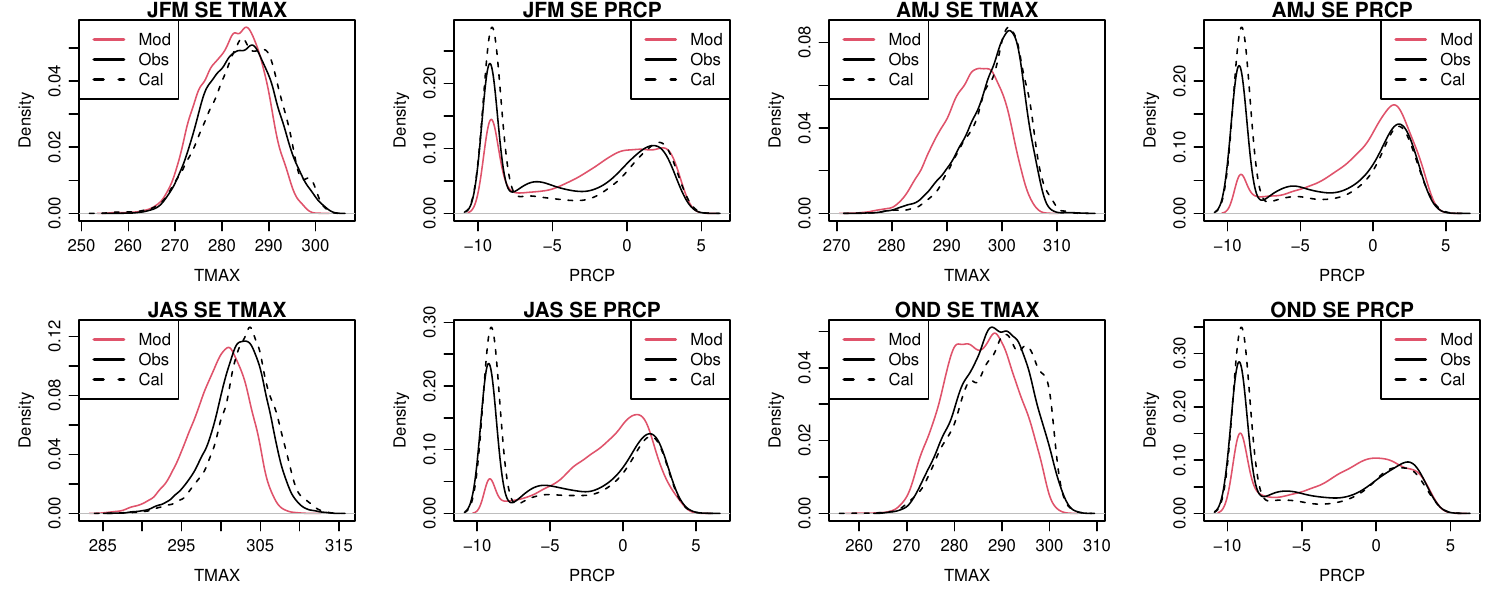}
         \caption{Seasonal density plots for validation period of 2001--2014.}
         \label{fig:density_seasonal_validation}
     \end{subfigure}
        \caption{Seasonal density plots for TMAX and PRCP data in the Southwest (SW) and Southeast (SE).}
        \label{fig:density_seasonal}
\end{figure}

This section provides additional seasonal and monthly calibration metrics of PRCP and TMAX. 
Figure \ref{fig:density_seasonal} plots seasonal densities of the two variables for both regions, for the training and validation periods. The seasons are defined as Spring (Jan--Mar), Summer (Apr--Jun), Fall (Jul--Sep), and Winter (Oct--Dec). The density of TMAX is well calibrated for both the training and validation periods. The PRCP densities are also well calibrated, though there are a few deviations around the extremes for the validation period. In particular, the proportion of zeros is overestimated for both regions in several of the months.

\begin{figure}
     \centering
     \begin{subfigure}[b]{\textwidth}
         \centering
        \includegraphics[scale=0.35]{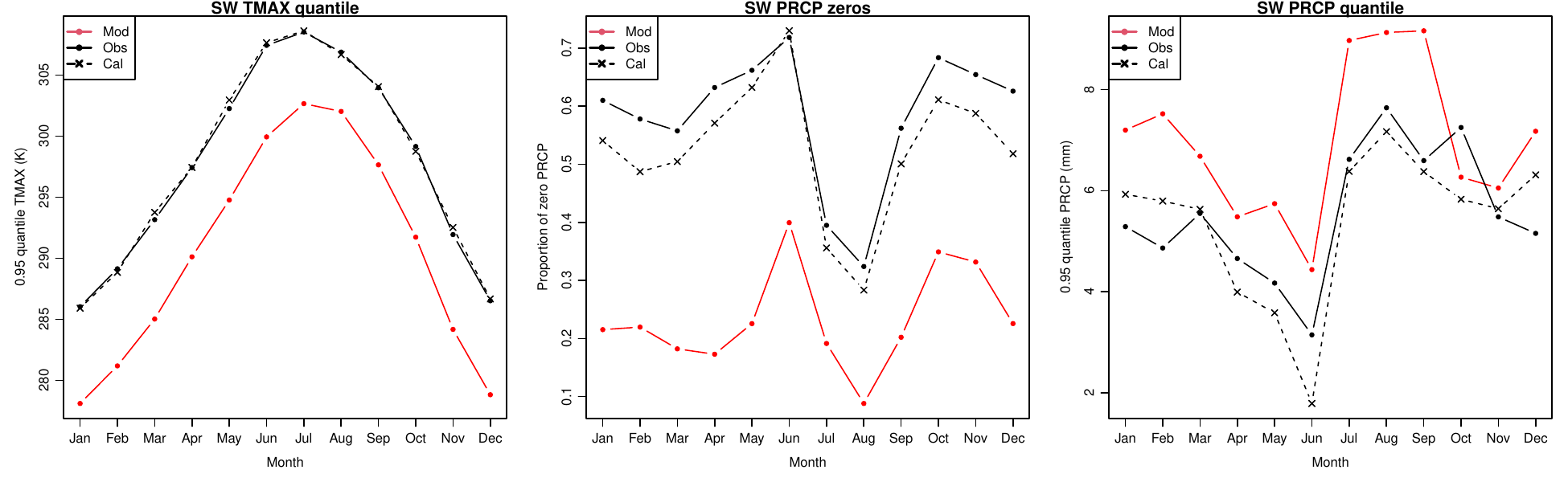}
         \includegraphics[scale=0.35]{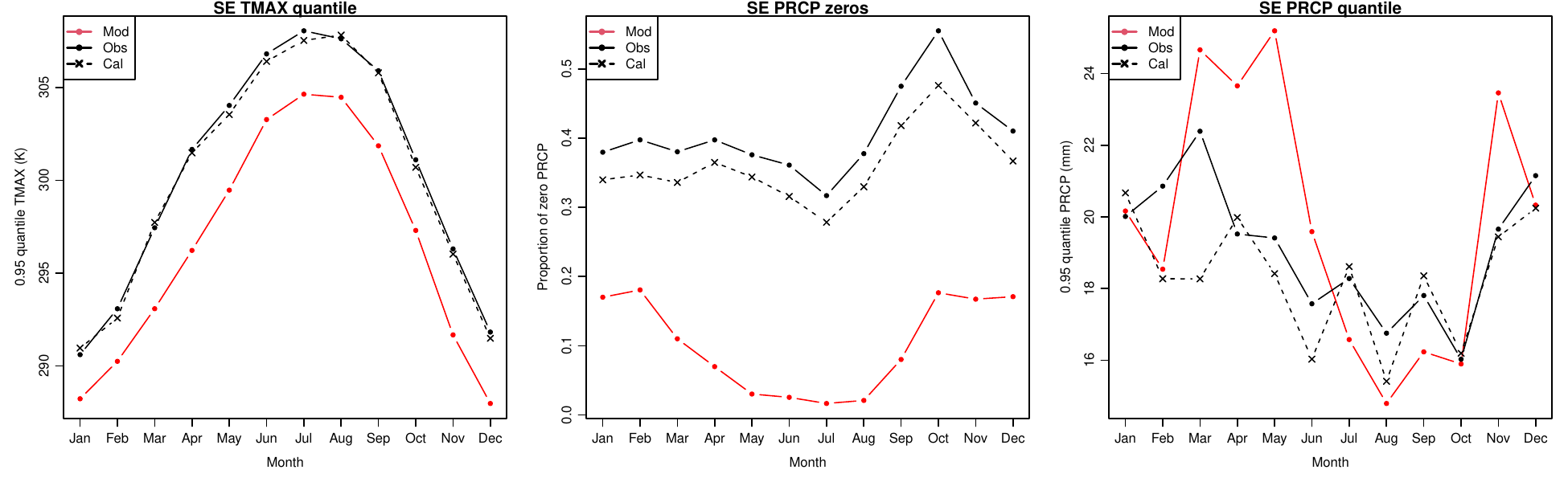}
         \caption{Upper quantiles and proportion of zeros for training period of 1951--2000.}
         \label{fig:quantiles_monthly_train}
     \end{subfigure}
     \vfill
     \begin{subfigure}[b]{\textwidth}
         \centering
         \includegraphics[scale=0.35]{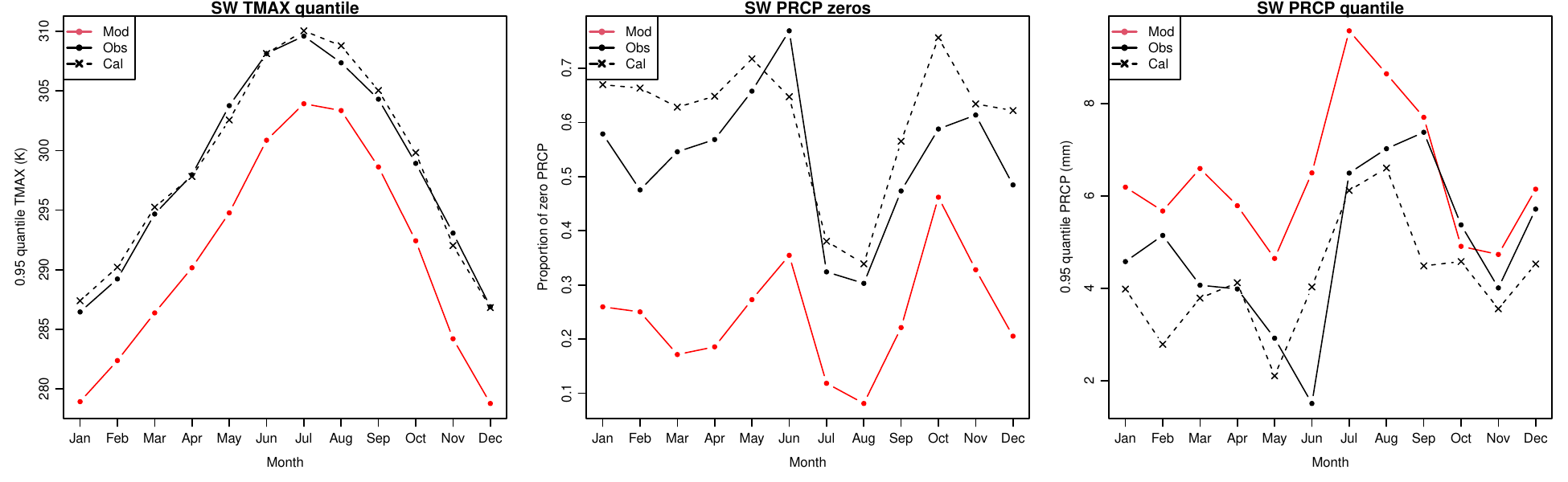}
         \includegraphics[scale=0.35]{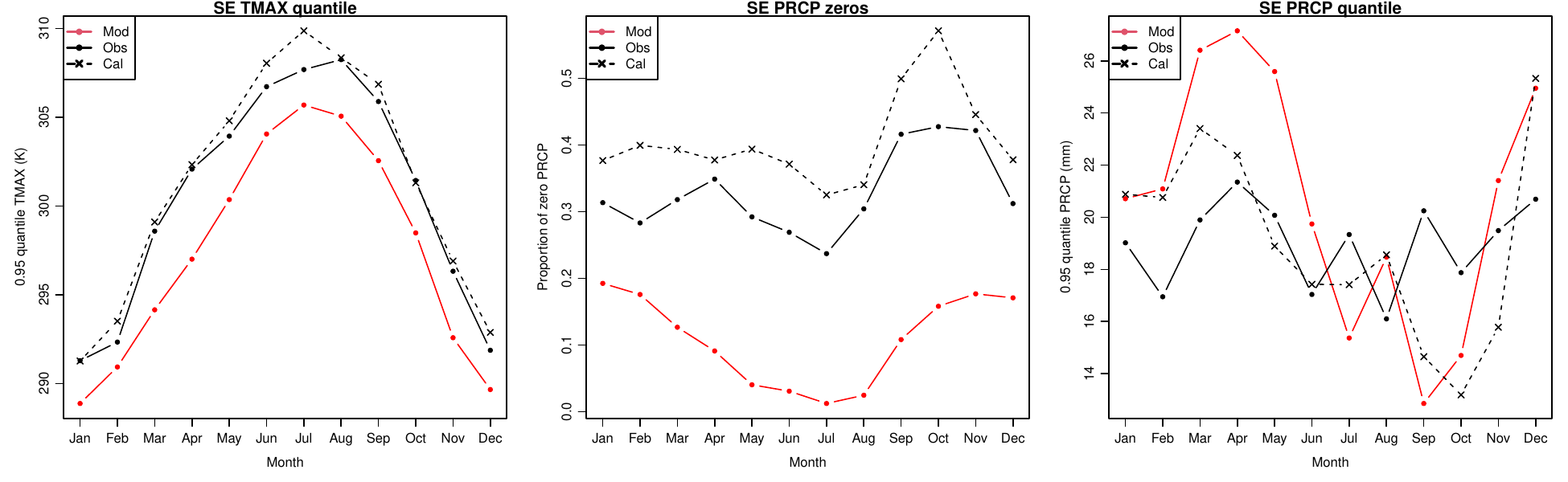}
         \caption{Upper quantiles and proportion of zeros for validation period of 2001--2014.}
         \label{fig:quantiles_monthly_train_validation}
     \end{subfigure}
        \caption{Monthly 0.95 quantiles (TMAX and PRCP) and proportion of zeros (PRCP) for the Southwest (SW) and Southeast (SE), pooled across all 25 locations.}
        \label{fig:quantiles_monthly}
\end{figure}

Figure \ref{fig:quantiles_monthly} plots the monthly 0.95 quantiles of both variables, and the proportion of zeros for PRCP; these are analogous to the monthly mean plots (Figure \ref{fig:means_monthly}) in the main text.  We note that the TMAX quantiles are well calibrated for both regions and both time periods (training and validation), with a little more error in the validation period compared to the training period. Similar to the corresponding plots for the means, the distribution of the quantiles are quite similar during the two time periods, and therefore are calibrated quite well. The proportion of zeros for the training period are slightly underestimated in both regions, since there is a massive difference between the proportions in the model and observational data; despite this, they are overall well calibrated. For the validation period, several of the months have lower proportion of zeros in the observational data compared to the historical period. In such situations, the calibrated data can overestimate the proportion. However, the calibration is in general adequate. Finally, for the upper quantiles of PRCP, the calibration for the SW is adequate for both the training and validation periods. However, for the SE, SPECD struggles to calibrate the quantiles for both the training and validation periods. This is similar behavior to their monthly means, and demonstrates an area where SPECD can be improved upon.

\end{appendix}
\end{document}